\documentclass[submission, Phys]{SciPost}
\usepackage{graphicx,amsmath,amssymb,epstopdf,wasysym}
\usepackage[space]{grffile}
\usepackage{hyperref}
\usepackage{mathtools}

\newcommand{\be}{\begin{equation}}
\newcommand{\ee}{\end{equation}}
\newcommand{\bea}{\begin{eqnarray}}
\newcommand{\eea}{\end{eqnarray}}
\newcommand{\ba}{\begin{eqnarray*}}
\newcommand{\ea}{\end{eqnarray*}}
\newcommand{\dagga}{{\phantom{\dagger}}}

\newcommand{\bq}{\mathbf{q}}

\newcommand{\bk}{\mathbf{k}}

\newcommand{\bkp}{\mathbf{k'}}

\newcommand{\bp}{\mathbf{p}}

\newcommand{\Rea}{\text{Re}}
\newcommand{\dis}{\displaystyle}

\newcommand{\fract}[2]{\frac{\dis #1}{\dis #2}}
\newcommand{\Tr}{\mathrm{Tr}}
\newcommand{\eqn}[1]{(\ref{#1})}

\newcommand{\bra}[1]{\langle #1 \mid}
\newcommand{\ket}[1]{\mid #1 \rangle}
\newcommand{\Bc}[1]{\,\,\,\hat{\!\!\!\bd{#1}}}
\newenvironment{eqs}%
{\begin{equation} \begin{aligned}}%
{\end{aligned} \end{equation} }
\newcommand{\beal}{\begin{eqs}}
\newcommand{\eal}{\end{eqs}}
\newcommand{\bealn}{\beal\nonumber}
\newcommand{\bw}{\begin{widetext}}
\newcommand{\ew}{\end{widetext}}
\newcommand{\esp}[1]{\text{e}^{#1}}

\newcommand{\ep}{\epsilon}
\newcommand{\bd}[1]{{\boldsymbol{#1}}}

\newcommand{\bnabla}{\bd{\nabla}}
\newcommand{\bkappa}{\bd{\kappa}}

\begin{document}

\begin{center}{\Large \textbf{Fermi-liquid corrections to the intrinsic anomalous Hall conductivity of topological metals }}\end{center}

\begin{center}
Ivan Pasqua\textsuperscript{1*},
Michele Fabrizio\textsuperscript{1}
\end{center}

\begin{center}
{\bf 1} International School for Advanced Studies (SISSA), Via Bonomea 265, I-34136 Trieste, Italy
\\
* ipasqua@sissa.it
\end{center}

\begin{center}
\today
\end{center}


\section*{Abstract}
{\bf
We show that topological metals lacking time-reversal symmetry have an
intrinsic non-quantised component of the anomalous Hall conductivity which is 
contributed not only by the Berry phase of quasiparticles on the Fermi surface, 
but also by Fermi-liquid corrections due to the residual interactions among quasiparticles, the Landau $f$-parameters. These corrections pair up with those 
that modify the optical mass with respect to the quasiparticle effective one, or the charge compressibility with respect to the quasiparticle density of states. 
Our result supports recent claims that the correct expressions for topological observables include vertex corrections besides the topological invariants built just upon the Green's functions. Furthermore, it demonstrates that such corrections are naturally accounted for by Landau's Fermi liquid theory, 
here extended to the case in which coherence effects between bands crossing the chemical potential and those that are instead away from it may play a crucial role, as in the anomalous Hall conductivity,  
and have important implications when those metals are on the verge of a doping-driven Mott transition, as we discuss. }

\vspace{10pt}
\noindent\rule{\textwidth}{1pt}
\tableofcontents\thispagestyle{fancy}
\noindent\rule{\textwidth}{1pt}
\vspace{10pt}

\section{Introduction}
\label{sec:intro}
In 2004, Haldane showed\cite{Haldane-PRL2004}, elaborating on earlier works   \cite{PhysRev.95.1154,PhysRev.112.739,PhysRevLett.88.207208,doi:10.1143/JPSJ.71.19}, that in metals where time-reversal symmetry is broken Landau's Fermi liquid theory \cite{Landau1} must be supplemented by a new, topological ingredient: the quasiparticle Berry phase. Specifically, he demonstrated  \cite{Haldane-PRL2004} that the expression for the non-quantised component of the intrinsic anomalous Hall conductivity, e.g., in ferromagnetic metals \cite{Review-Nagaosa-MacDonald}, can be recast as an integral over the quasiparticle Fermi surface, as later confirmed in 
\textit{ab initio} calculations \cite{PhysRevB.76.195109}, and in agreement with the spirit of Landau's Fermi liquid theory \cite{Landau1}. \\
However, Haldane's 
work leaves open an important issue that we now discuss. Landau's energy functional 
for quasiparticles \cite{Landau1} includes a quasiparticle energy, which we hereafter denote 
as the Hamiltonian matrix $\hat{H}_*(\bk)$ having in mind a multiband Fermi liquid, and a residual 
interaction among quasiparticles, defined in terms of the so-called Landau $f$-parameters that we write as a tensor $\hat{f}(\bk,\bkp)$ in band space. It is tempting to assume that the quasiparticle topology is encoded just in  
the quasiparticle Hamiltonian $\hat{H}_*(\bk)$ and its Bloch eigenstates 
$\ket{\!\!\psi_n(\bk)}$, a relation that was already put forth by 
Haldane \cite{Haldane-PRL2004}.  
Nonetheless, there is evidence that such reasonable choice may not be correct \cite{CHEN2017345,Andrea-PRB2023,PhysRevLett.131.236601,PRB-Cenke,PRL-Guerci,PRL-Konig,arXiv-Martin,PRB-Principi}, 
namely that the Berry curvature calculated through the quasiparticle Bloch states $\ket{\!\!\psi_n(\bk)}$ not necessarily yields the anomalous Hall conductivity. For instance, the authors of Ref.~\cite{CHEN2017345} compute the electromagnetic response of a Fermi liquid endowed with a finite Berry curvature through a linearised kinetic equation. They found corrections to the anomalous Hall coefficient, which they identify as coming from an electric dipole moment of the quasiparticles. \\
It is legitimate to wonder whether this evidence is still compatible with 
the formal, microscopic derivation of Landau's Fermi liquid theory \cite{Nozieres&Luttinger-1}, namely, if the latter does account for 
Fermi liquid corrections to the anomalous Hall conductivity. This is precisely the aim of the present study. \\
The structure of the paper is the following. In Sec. \ref{Hall conductivity within Landau's Fermi liquid theory}  we introduce the multi-band formalism of Landau's Fermi-liquid theory and we derive the Hall conductivity tensor, including the Fermi liquid corrections due to the residual interactions between the quasiparticles. In Sec. \ref{A toy model calculation} we specialize our discussion to a model topological metal with broken time-reversal symmetry, where we can explicitly verify that the corrections to the anomalous Hall conductivity stem solely from the quasiparticle Fermi surface. In Sec. \ref{Conclusions} we explore the implications of these Fermi liquid corrections in the computation of the topological observables, specifically in strongly correlated topological metals on the verge of a doping-driven Mott transition. The appendices are instead devoted to a more formal derivation of Landau's Fermi liquid theory in a multi-band system. 

\section{Hall conductivity within Landau's Fermi liquid theory}
\label{Hall conductivity within Landau's Fermi liquid theory}

We assume a periodic system of interacting 
electrons, thus in absence of impurities and their extrinsic contributions to the 
anomalous Hall effect \cite{Review-Nagaosa-MacDonald}, at sufficiently low temperature to safely discard any quasiparticle decay rate, and, as mentioned, we consider the general case of many bands crossing the chemical potential that can be described within Landau's Fermi-liquid theory. However, we shall deal with this theory following an early observation by Leggett \cite{RevModPhys.47.331}. Indeed, one may realise \cite{RevModPhys.47.331,libro} that Landau's Fermi-liquid theory basically shows \cite{Nozieres&Luttinger-1} that the low-frequency, long-wavelength and low-temperature response functions of the physical electrons correspond 
to those of a system of interacting quasiparticles treated by the Hartree-Fock (HF)  
plus the random phase (RPA) approximations, apart from important caveats that we discuss in Appendix \ref{Appendix}. In Appendix \ref{Formal} we present a detailed derivation of 
Leggett's result \cite{RevModPhys.47.331} generalised to the case of a multi-band Fermi liquid. 
\\
Therefore, let us consider quasiparticles described by an interacting Hamiltonian, hereafter setting $\hbar=1$, 
\beal
H_\text{qp} &= H_{0} + H_\text{int}\,,
\label{Hqp}
\eal
where 
\bealn
H_{0}=\sum_\bk\,\sum_{\alpha\beta}\, c^\dagger_{\alpha\bk}\, H_{0\,\alpha\beta}(\bk)\,
c^\dagga_{\beta\bk}\,,
\eal
is a one-body term, e.g., a tight binding Hamiltonian, with $c^\dagga_{\alpha\bk}$ the annihilation operators of the quasiparticles (here $\alpha$ represents both the band index and the spin $\sigma$), and $H_\text{int}$ the interaction, which we do not even need to specify.\\
Within HF, one replaces the interacting Hamiltonian \eqn{Hqp} with the non-interacting one  
\beal
\hat{H}_*(\bk) = \hat{H}_{0}(\bk) + \hat{\Sigma}_\text{HF}\big[\bk,\hat{G}_*\big]\,,
\label{self-consistency 1}
\eal
which is just the quasiparticle Hamiltonian we mentioned above. In \eqn{self-consistency 1}, 
$\hat{H}_{0}(\bk)$ is the matrix with elements $H_{0\,\alpha\beta}(\bk)$, 
$\hat{\Sigma}_\text{HF}\big[\bk,\hat{G}_*\big]$ includes only the Hartree and Fock diagrams 
of the self-energy functional, and 
\beal
\hat{G}_*(i\ep,\bk) &= \fract{1}{\;i\ep- \hat{H}_*(\bk)\;}\;,
\label{self-consistency 2}
\eal
is the HF Green's function in Matsubara frequencies. RPA is a symmetry-conserving scheme consistent with HF \cite{libro}, which amounts to calculate response functions using the Green's function \eqn{self-consistency 2} and the irreducible scattering vertex in the particle-hole channel defined as 
\beal
\fract{\delta \hat{\Sigma}_\text{HF}\big[\bk,\hat{G}_*\big]}{\delta \hat{G}_*(i\ep,\bkp)} = 
\hat{f}(\bk,\bkp)\,,
\label{self-consistency 3}
\eal
thus providing the definition of the Landau $f$-parameter tensor. 
Accordingly, the quasiparticle energies $\ep_\ell(\bk)$ and eigenfunctions
$\ket{\!\psi_\ell(\bk)}$, using roman letters to distinguish them from the greek ones that label the 
original basis, are obtained by diagonalising  
\beal
H_{* \alpha\beta}(\bk) &= H_{0 \alpha\beta}(\bk) + 
\fract{1}{V}\sum_{\bkp\mu\nu}\, f_{\alpha\nu,\mu \beta}(\bk,\bkp)\, 
\langle\,c^\dagger_{\nu\bkp}\,c^\dagga_{\mu \bkp}\,\rangle \\
&= H_{0 \alpha\beta}(\bk) + 
\fract{T}{V}\sum_{\bkp\mu\nu}\sum_\ep\,\esp{i\ep 0^+}\,f_{\alpha\nu,\mu \beta}(\bk,\bkp)\, G_{*\,\mu\nu}(i\ep,\bkp)\,,
\label{H* original basis}
\eal
where $V$ is the volume, $f_{\alpha \gamma,\delta \beta}(\bk,\bkp)$ are, through \eqn{self-consistency 3}, the components of 
$\hat{f}(\bk,\bkp)$,     
and the expectation values, to be determined self-consistently, are over the HF ground state. 
The diagonalization is accomplished by a unitary transformation $\hat{U}(\bk)$, 
\bealn
\hat{U}(\bk)^\dagger\,\hat{H}_*(\bk)\,\hat{U}(\bk) = \hat{\ep}_*(\bk)\,,
\eal
where $\hat{\ep}_*(\bk)$ is the diagonal matrix with elements $\ep_\ell(\bk)$.\\ 

\noindent
By means of the Ward-Takahashi identities, Nozi\'eres and Luttinger \cite{Nozieres&Luttinger-1} showed that in the so-called static or $q$-limit, the transferred frequency vanishing before the transferred momentum, the fully-interacting physical current vertex multiplied by the quasiparticle residue, 
which we hereafter denote as $\Bc{J}{\,^q}(\bk)$, corresponds to $\bnabla_\bk \,\hat{H}_*(\bk)$ for the quasiparticles. In the diagonal basis, which we define as 
HF basis, that correspondence reads 
\beal
\Bc{J}{\,^q}(\bk) &\equiv\hat{U}(\bk)^\dagger\,\bnabla_\bk\,\hat{H}_*(\bk)\,\hat{U}(\bk) 
= 
\hat{U}(\bk)^\dagger\,\bnabla_\bk\,\Big(\hat{U}(\bk)\,\hat{\ep}_*(\bk)\,\hat{U}(\bk)^\dagger\Big)\,\hat{U}(\bk)\\
&=
\bnabla_\bk\, \hat{\ep}_*(\bk)  - \Big[\hat{\ep}_*(\bk)\,,\, \hat{U}(\bk)^\dagger\,\bnabla_\bk\,\hat{U}(\bk)\Big]\equiv \bnabla_\bk\,\hat{\ep}_*(\bk) 
+ i\,\Big[\hat{\ep}_*(\bk),\Bc{\mathcal{A}}^0(\bk)\Big]\,,
\label{current vertex zero}
\eal
where the first term is diagonal and the second off-diagonal in the band indices.  
Here, the hermitian operator 
$\Bc{\mathcal{A}}^0(\bk) =i\,\hat{U}(\bk)^\dagger\,\bnabla_\bk\,\hat{U}(\bk)$ 
is the bare Berry connection of the quasiparticles, whose matrix elements are also equal to 
\beal
\bd{\mathcal{A}}^0_{\ell m}(\bk) & = i\, \bra{\psi_\ell(\bk)}
\bnabla_\bk\,\psi_m(\bk)\rangle = -i\,  \bra{\bnabla_\bk\,\psi_\ell(\bk)}
\psi_m(\bk)\rangle \,,
\label{bare Berry connection}
\eal
through which the diagonal part of the bare Berry curvature is $\bd{\Omega}^0_{\ell\ell}(\bk) = -\bnabla_\bk\times\bd{\mathcal{A}}_{\ell\ell}^0(\bk)$. \\

\noindent
The reason why we have reformulated Landau's Fermi-liquid theory in such more complex language 
is to make it clear that quasiparticles are interacting, though interaction is treated within HF plus RPA. 
This point becomes important when one wants to compute topological observables. As already mentioned, the first temptation is to use the eigenfunctions 
$\ket{\!\psi_\ell(\bk)}$ and calculate the topological invariants as one would do for non-interacting 
electrons, see \eqn{bare Berry connection}. However, this procedure might be incorrect just because the 
quasiparticles do interact with each other. \\
A more rigorous approach is to directly calculate the anomalous Hall conductivity as dictated by Landau's Fermi liquid theory \cite{Nozieres&Luttinger-1}. In  Appendix \ref{Appendix} we derive the expression of the matrix elements 
$\sigma^\text{H}_{ab}=-\sigma^\text{H}_{ba}$, $a\not=b$ labelling the spatial components, of the Hall conductivity tensor, see \eqn{Hall 3}, 
which read, explicitly, 
\beal
\sigma_{ab}^\text{H} &=- i\,\fract{e^2}{V}\,\sum_{\bk}\, \sum_{\ell\not= m}\,
J^\omega_{a\,\ell m}(\bk)\; \fract{\;f\big(\ep_\ell(\bk)\big) - f\big(\ep_m(\bk)\big)\;}
{\;\big(\ep_m(\bk)-\ep_\ell(\bk)\big)^2\;}\;
J^\omega_{b\,m\ell}(\bk)\\
&= - i\,\fract{e^2}{V}\,\sum_{\bk}\,\sum_{\ell\not= m}\, f\big(\ep_\ell(\bk)\big)\, 
\fract{\;J^\omega_{a\,\ell m}(\bk)\,J^\omega_{b\,m\ell}(\bk) - J^\omega_{b\,\ell m}(\bk)\,J^\omega_{a\,m \ell }(\bk)\;}
{\;\big(\ep_m(\bk)-\ep_\ell(\bk)\big)^2\;}\\
&\equiv \fract{e^2}{V}\,\sum_{\bk\ell}\, f\big(\ep_\ell(\bk)\big)\, \ep_{abc}\,\Omega_{c\,\ell\ell}(\bk)
\,,
\label{Hall conductivity 1}
\eal
where $f(x)$ is the Fermi distribution function, and 
\beal
\bd{\Omega}_{\ell\ell}(\bk) &= -i\,\sum_{m\not=\ell}\, \fract{\;\bd{J}^\omega_{\ell m}(\bk)\times \bd{J}^\omega_{m\ell}(\bk) \;}
{\;\big(\ep_m(\bk)-\ep_\ell(\bk)\big)^2\;}\;,
\label{quasiparticle Omega}
\eal
is therefore the true quasiparticle Berry curvature. The dynamic,  
$\omega$-limit of the current vertex, the opposite of the $q$-limit, is related to \eqn{current vertex zero} through \eqn{Hall 4}, specifically, 
\beal
\bd{J}^\omega_{\ell m}(\bk) &= \bd{J}^q_{\ell m}(\bk)- \fract{1}{V}\,\sum_{\bkp n}\,\Gamma^\omega_{\ell n, n m}(\bkp,\bk)\,
\bnabla_\bkp\,f\big(\ep_n(\bkp)\big)\,,
 \label{current vertex}
\eal
where $\Gamma^\omega$ is the $\omega$-limit of the reducible scattering vertex \eqn{Gamma}. 
The term in \eqn{quasiparticle Omega} with the bare current vertices, i.e., 
$\Gamma^\omega=0$ in \eqn{current vertex},  
reproduces the bare Berry curvature $\bd{\Omega}^0_{\ell\ell}(\bk)$, while the additional terms correspond to the desired Fermi liquid corrections, and indeed derive, see the second term in \eqn{current vertex}, from the quasiparticle Fermi surface. It is therefore reasonable to argue that also the corrections to the anomalous Hall conductivity come from the quasiparticle Fermi surface, though this conclusion seems not immediately obvious looking at \eqn{Hall conductivity 1} and \eqn{current vertex}. Indeed, it may seem counterintuitive that a quantity evaluated within the Landau's theory of Fermi liquids could have contributions from bands that do not cross the chemical potential. However, we believe such a paradox to be merely apparent.  
In the next section, we analyze a specific example and show explicitly 
that the above surmise is true.\\ 
Nonetheless, we emphasize that, while the corrections do derive from quasiparticles at the Fermi surface, equation \eqn{Hall conductivity 1} accounts also for cases where there still are occupied bands that contribute to the anomalous Hall conductivity. It might at first looks odd that the contribution of an occupied band may not be quantized because of the interaction with bands crossing the chemical potential, but that does not contradict any physical principle as long as the contribution 
becomes again quantized when no band cross anymore the chemical potential, which is clearly the case here, see \eqn{current vertex}.

\section{A toy model calculation}
\label{A toy model calculation}

We consider the Bernevig, Hughes and Zhang (BHZ) model \cite{BHZ} for a quantum spin-Hall insulator on a square lattice. In particular, we assume the model with full spin polarisation, 
only spin up bands being occupied, and at density $n=1+\delta$, in which case the model 
describes a topological metal with broken time-reversal symmetry.\\
The Hamiltonian for the spin-up quasiparticles is assumed to be    
\beal
\hat{H}_*(\bk) &= \big(\ep(\bk)-\mu\big)\,\sigma_0 + \big(M-t(\bk)\big)\,\sigma_3 
+\lambda(\bk)\,\sin k_x\,\sigma_1 -\lambda(\bk)\,\sin k_y\,\sigma_2 \\
&= \big(\ep(\bk)-\mu\big)\,\sigma_0 + 
\bd{B}(\bk)\cdot\bd{\sigma}\,,
\label{Ham}
\eal
where $\mu$ crosses either the valence or the conduction bands, 
the identity, $\sigma_0$, and the Pauli matrices $\sigma_a$, $a=1,2,3$, act on the orbital space, and 
\beal
\bd{B}(\bk) &= \big(\lambda(\bk)\,\sin k_x, -\lambda(\bk)\,\sin k_y, M-t(\bk)\big)\\
&= B(\bk)\,\Big(\cos\phi(\bk)\,\sin\theta(\bk),
\sin\phi(\bk)\,\sin\theta(\bk),\cos\theta(\bk)\Big)\\
&= B(\bk)\,\Big(
\sin\theta(\bk)\,\bd{v}_2(\bk)
+ \cos\theta(\bk)\,\bd{v}_3(\bk)\Big)\,.
\label{B(k)}
\eal
Here, $\bd{v}_2(\bk)=(\cos\phi(\bk),\sin\phi(\bk),0)$ and $\bd{v}_3(\bk)=(0,0,1)$
are orthogonal unit vectors that form a basis together with $\bd{v}_1(\bk)=\bd{v}_2(\bk)\times\bd{v}_3(\bk)$. This, in turn, implies that $\sigma_0$ and 
$\bd{v}_a(\bk)\cdot\bd{\sigma}$, $a=1,2,3$, form a basis of $2\times 2$ matrices. To simplify the notations, in what follows we use the definition $\bd{v}_0(\bk)\cdot\bd{\sigma}\coloneqq = \sigma_0$. \\
The quasiparticle Hamiltonian 
\eqn{Ham} is assumed to be invariant under inversion $\mathcal{I}$, $\bk\to-\bk$ and $\sigma_a\to-\sigma_a$ for $a=1,2$, 
and fourfold rotations $C_4$, $k_x\to k_y\,\land\,k_y\to - k_x$ and $\sigma_1\to-\sigma_2\,\land\,\sigma_2\to\sigma_1$. This requires that the parameters $\ep(\bk)$, $t(\bk)$ and $\lambda(\bk)$ in \eqn{Ham} are invariant under both $\mathcal{I}$ and $C_4$. In addition, in order to clearly distinguish $\mu$ and $M$ from, respectively, $\ep(\bk)$ and $t(\bk)$, we assume that the latter average out at zero over the Brillouin zone.\\
The Hamiltonian $\hat{H}_*(\bk)$ is diagonalised by the unitary transformation 
\bealn
\hat{U}(\bk) = \esp{ i\frac{\theta(\bk)}{2}\,\bd{v}_1(\bk)\cdot\bd{\sigma}} = 
\cos\fract{\theta(\bk)}{2} + i\,\sin\fract{\theta(\bk)}{2}\;\bd{v}_1(\bk)\cdot\bd{\sigma}\,,
\eal 
namely $\hat{U}(\bk)^\dagger\,\hat{H}_*(\bk)\,\hat{U}(\bk) = \left(\ep(\bk)-\mu\right)\,\sigma_0 + B(\bk)\,\sigma_3$. 
\begin{figure}
\centerline{\includegraphics[width=0.8\textwidth]{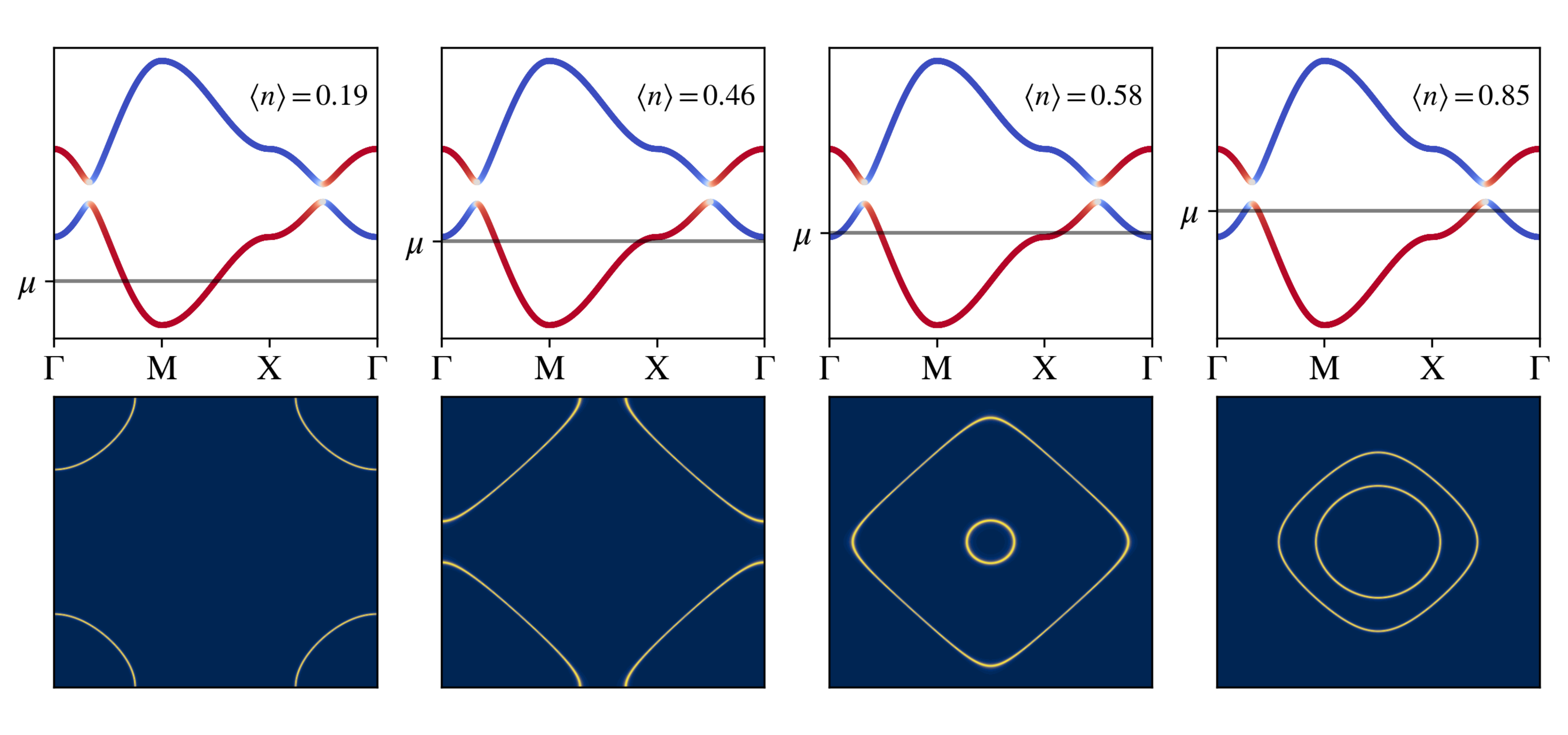}}

\centerline{\includegraphics[width=0.8\textwidth]{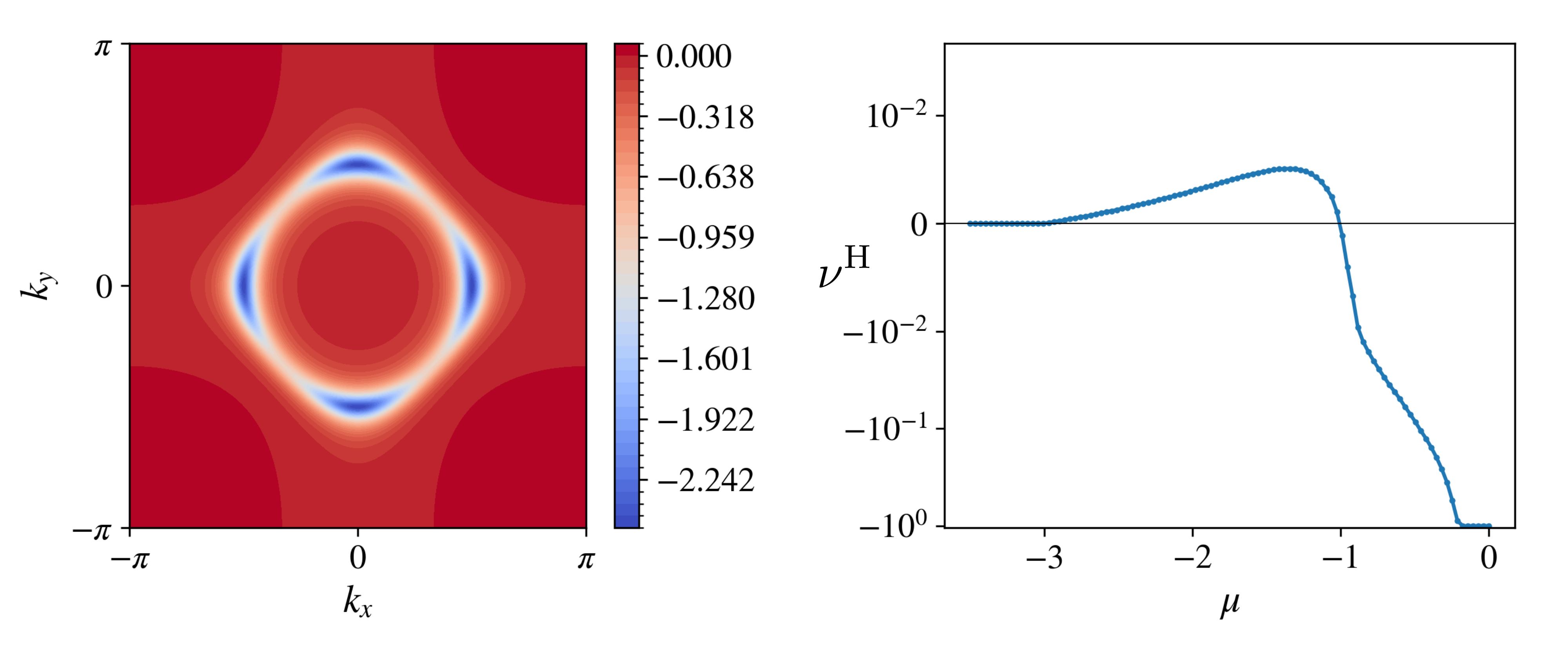}}
\caption{Top panel: band structures of \eqn{Ham} with $\ep(\bk)=0$, $M=1$, 
$t(\bk)=\cos k_x+\cos k_y$ and $\lambda(\bk)=0.2$ 
(top), and Fermi surfaces (bottom) for different values of $\mu$ corresponding to  hole doping. Bottom panel: distribution in momentum space of the bare Berry 
curvature $\Omega^0(\bk)$ of the valence band (left); integral $\nu^\text{H}$ 
of the bare Berry 
curvature in units of $2\pi$ over the occupied Fermi volume (right) as function of $\mu<0$. When $2|\mu|$ is smaller than the gap, $\nu^\text{H}=-1$ keeps its 
quantised value, otherwise it deviates reaching zero when the valence band empties, $\mu=-3$. In reality, $\nu^\text{H}$ crosses zero also when the Fermi surface changes character, from hole-like to 
electron-like. We also note that $\Omega^0(\bk)$ of the valence band 
is peaked in magnitude at the top of the band, when the orbital character changes. This explain the fast decrease in magnitude upon hole doping.}
\label{Fig1}
\end{figure}
We readily find that 
\beal
\Bc{\mathcal{A}}^0(\bk)&=i\,\hat{U}(\bk)^\dagger\,\bnabla_\bk\,\hat{U}(\bk) =  
-\fract{\;\bnabla_\bk\,\theta(\bk)}{2}\,\bd{v}_1(\bk)\cdot\bd{\sigma}\\
&\qquad 
-\fract{\;\bnabla_\bk\,\phi(\bk)}{2}\,\sin\theta(\bk)\,\bd{v}_2(\bk)\cdot\bd{\sigma}  -\bnabla_\bk\,\phi(\bk)\,\sin^2\fract{\theta(\bk)}{2}\;\bd{v}_3(\bk)\cdot\bd{\sigma}\\
&=  \sum_{a=1}^3\, \bd{\mathcal{A}}^0_a(\bk)\,\bd{v}_a(\bk)\cdot\bd{\sigma}\,.
\label{Berry connection BHZ}
\eal
We can make a similar expansion for the currents, 
\bealn
\Bc{J}^q(\bk) &= 
\sum_{a=0}^3\, \bd{J}^q_a(\bk)\,\bd{v}_a(\bk)\cdot\bd{\sigma}\,,&
\Bc{J}^\omega(\bk) &= 
\sum_{a=0}^3\, \bd{J}^\omega_{a}(\bk)\,\bd{v}_a(\bk)\cdot\bd{\sigma}\,,
\eal
where $\bd{J}^q_0(\bk) = \bnabla_\bk\,\ep(\bk)$, $\bd{J}^q_3(\bk) = \bnabla_\bk\,B(\bk)$ and 
$\bd{J}^q_a(\bk) = 2\,B(\bk)\,\ep_{ab3}\,\bd{\mathcal{A}}^0_b(\bk)$ for 
$a=1,2$. In the top panel of Fig.~\ref{Fig1} we show the band structure of \eqn{Ham} at $\ep(\bk)=0$, $M=1$, 
$t(\bk)=\cos k_x+\cos k_y$ and $\lambda(\bk)=0.2$, indicating the chemical potential $\mu$ for the different hole doping levels and the corresponding Fermi surfaces. In the bottom panel, we draw the momentum distribution of the bare Berry curvature of the valence band, as well as its 
integral in units of $2\pi$ over the occupied Fermi volume as function of the chemical potential $\mu<0$.\\

\noindent
We note that $\bd{v}_a(\bk)\cdot\bd{\sigma}$, $a=0,\dots,3$, are invariant under both $\mathcal{I}$ 
and $C_4$, and $\bnabla_\bk\,\phi(\bk)$ as well as 
$\bnabla_\bk\,\theta(\bk)$ transform as conventional vectors. Since the $f$-parameters contribute to 
the quasiparticle Hamiltonian \eqn{Ham} through the mean-field decoupling, the expression of 
$\bd{B}(\bk)$ in \eqn{B(k)} implies that the most general $\hat{f}$-tensor in the original basis 
must be of the form
\beal
\hat{f}(\bk,\bkp) &= \sum_{a,b=0,2,3}\,f_{ab}(\bk,\bkp)\,
\big(\bd{v}_a(\bk)\cdot\bd{\sigma}\big)\otimes
\big(\bd{v}_b(\bkp)\cdot\bd{\sigma}\big)\,,
\label{f-tensor}
\eal
thus lacking terms that involve $\bd{v}_1(\bk)\cdot\bd{\sigma}$,  
and where symmetry only requires that the expansion coefficients are 
invariant under $\mathcal{I}$ and $C_4$. Upon rotation in the HF basis, 
\eqn{f-tensor} maintains the same form with modified coefficients, which we still 
denote as $f_{ab}(\bk,\bkp)$ for sake of simplicity. We can 
further identify two distinct tensors, $\hat{f}^0(\bk,\bkp)=\hat{f}^0(-\bk,\bkp)=\hat{f}^0(\bk,-\bkp)$ and $\hat{f}^1(\bk,\bkp)=-\hat{f}^1(-\bk,\bkp)=-\hat{f}^1(\bk,-\bkp)$, using the same notation as in conventional single-band Fermi liquids \cite{Pines&Nozieres}. It is precisely $\hat{f}^1(\bk,\bkp)$ that 
may contribute to \eqn{current vertex}, which, without loss of generality, can be taken in the HF basis of the form 
\bealn
\hat{f}^1(\bk,\bkp) &= 
\sum_{a,b=0,2,3}\,f^1_{ab}(\bk,\bkp)\,\bd{J}^q_a(\bk)\cdot
\bd{J}^q_b(\bkp)\,\big(\bd{v}_a(\bk)\cdot\bd{\sigma}\big)\otimes
\big(\bd{v}_b(\bkp)\cdot\bd{\sigma}\big)\,,
\eal
where, for consistency, $f^1_{ab}(\bk,\bkp)=f^1_{ab}(-\bk,\bkp)=f^1_{ab}(\bk,-\bkp)$ vary in momentum space orthogonally 
to the corresponding currents, i.e., 
\beal
\bd{J}^q_a(\bk)\cdot\bnabla_\bk\,f^1_{ab}(\bk,\bkp) &= 
\bnabla_\bkp\,f^1_{ab}(\bk,\bkp)\cdot \bd{J}^q_b(\bkp) =0\,,
\label{constraints}
\eal
for $a,b=0,2,3$.
In reality, 
only the components $f^1_{ab}(\bk,\bkp)$, $a=0,2,3$ and $b=0,3$, contribute to 
\eqn{current vertex}, in which case \eqn{Gamma} implies that $\Gamma^{1\omega}_{ab}(\bk,\bkp)\equiv f^1_{ab}(\bk,\bkp)$. 
It is worth remarking that $\hat{f}(\bk,\bkp)$ in \eqn{f-tensor}
contributes to $\hat{H}_*(\bk)$ in \eqn{Ham} through the HF self-energy, 
see \eqn{H* original basis}. Therefore, if we reasonably assume that the 
HF ground state does not carry any current, then the mean-field terms 
generated by $\hat{f}^1(\bk,\bkp)$ must vanish at self-consistency, which 
requires, at the very least, non-singular coefficients $f^1_{ab}(\bk,\bkp)$. This observation will be useful later on.  \\
We further define the diagonal matrix 
$\hat{\mathcal{P}}(\bk)$ with elements $f\big(\ep_\ell(\bk)\big)$, 
which we can write as 
\bealn 
\hat{\mathcal{P}}(\bk) &= \fract{\;f\big(\ep_1(\bk)\big)+f\big(\ep_2(\bk)\big)\;}{2}\;\bd{v}_0(\bk)\cdot
\bd{\sigma}+ \fract{\;f\big(\ep_1(\bk)\big)-f\big(\ep_2(\bk)\big)\;}{2}\;\bd{v}_3(\bk)\cdot
\bd{\sigma}\\
&\equiv \mathcal{P}_0(\bk)\,\bd{v}_0(\bk)\cdot
\bd{\sigma} + \mathcal{P}_3(\bk)\,\bd{v}_3(\bk)\cdot
\bd{\sigma}\,,
\eal
where the conduction band 1 corresponds to $\sigma_3=+1$ with energy $\ep_1(\bk) 
=\ep(\bk)-\mu+B(\bk)$, and the valence band 2 to $\sigma_3=-1$ and $\ep_2(\bk) 
=\ep(\bk)-\mu-B(\bk)$. With those definitions, \eqn{current vertex} transforms into an equation for each component in the matrix basis, specifically, and exploiting the spatial symmetries, 
\beal
\bd{J}^\omega_{a}(\bk) &=  \bd{J}^q_a(\bk)
- \big(1-\delta_{a,1}\big)\,\fract{2}{V}\,\sum_\bkp\,\sum_{b=0,3}\,f^1_{ab}(\bk,\bkp)\,\bnabla_\bkp\,
\mathcal{P}_b(\bkp)\,\Big(\bd{J}^q_a(\bk)\cdot
\bd{J}^q_b(\bkp)\Big)\\
&= \bd{J}^q_a(\bk)\,\Bigg\{ 1 - \big(1-\delta_{a,1}\big)\,\fract{1}{V}\,\sum_\bkp\,\sum_{b=0,3}\,
f^1_{ab}(\bk,\bkp)\,\bnabla_\bkp\,
\mathcal{P}_b(\bkp)\cdot\bd{J}^q_b(\bkp)\Bigg\}\\
&\equiv \bd{J}^q_a(\bk)\,\Bigg(1+\big(1-\delta_{a,1}\big)\,\fract{F^1_{a}(\bk)}{2}\Bigg)\,.
\label{current vertex BHZ}
\eal  
The anomalous Hall conductivity $\sigma^\text{H}=\sigma_{xy}^\text{H}$ in \eqn{Hall conductivity 1} can be simply written as 
\beal
\sigma^\text{H} &= -i\,\fract{e^2}{V}\,\sum_\bk\,\sum_{a,b=1}^2\,
\fract{\;J^\omega_{x a}(\bk)\, J^\omega_{y b}(\bk)\;}{\;4B(\bk)^2\;}\; 
\Tr\bigg(
\Big[\mathcal{P}_3(\bk)\,\bd{v}_3(\bk)\cdot\bd{\sigma},\bd{v}_a(\bk)\cdot\bd{\sigma}\Big]
\,\bd{v}_b(\bk)\cdot\bd{\sigma}\bigg)
\\
&=\fract{e^2}{2V}\,\sum_\bk\,\sum_{a,b=1}^2\,
\fract{\;\mathcal{P}_3(\bk)\;}{\;B(\bk)^2\;}\; \ep_{ab3}\,
\bd{J}^\omega_{a}(\bk)\times\bd{J}^\omega_{b}(\bk)
\cdot\bd{v}_3(\bk)\\
&=-\fract{e^2}{V}\,\sum_\bk\,\mathcal{P}_3(\bk)\,\bigg(1+\fract{F^1_{2}(\bk)}{2}\bigg)
\,
\bnabla_\bk\times\Big(\cos\theta(\bk)\,\bnabla_\bk\,\phi(\bk)\Big)
\cdot\bd{v}_3(\bk)\\
&= \fract{e^2}{V}\,\sum_\bk\,\cos\theta(\bk)\,\bigg(1+\fract{F^1_{2}(\bk)}{2}\bigg)\,
\bnabla_\bk\,\mathcal{P}_3(\bk)\times
\bnabla_\bk\,\phi(\bk)
\cdot\bd{v}_3(\bk)\\
&\qquad + \fract{e^2}{2V}\,\sum_\bk\,\cos\theta(\bk)\,\mathcal{P}_3(\bk) \,
\bnabla_\bk\,F^1_{2}(\bk)\times
\bnabla_\bk\,\phi(\bk)
\cdot\bd{v}_3(\bk)\,.
\label{Hall conductivity BHZ}
\eal
We note that the first term in the last equality of \eqn{Hall conductivity BHZ} is a genuine Fermi surface contribution. Concerning the last term, we recall that $\bd{J}_2^q(\bk)\propto 
\bd{\mathcal{A}}^0_1(\bk)\propto 
\bnabla_\bk\,\theta(\bk)$ and, because of \eqn{constraints}, the vector product 
$\bnabla_\bk\,F^1_{2}(\bk)\times\bnabla_\bk\,\phi(\bk)$ is proportional 
to $\sin k_x\,\sin k_y$, odd under $C_4$, which therefore averages out at zero upon summing over $\bk$ since 
$\mathcal{P}_3(\bk)$ and $\cos\theta(\bk)$ are both invariant. In conclusion 
\beal
\sigma^\text{H} &= \fract{e^2}{V}\,\sum_\bk\,\bigg(1+\fract{F^1_{2}(\bk)}{2}\bigg)\,\cos\theta(\bk)\; \bnabla_\bk\,\mathcal{P}_3(\bk)\times
\bnabla_\bk\,\phi(\bk)
\cdot\bd{v}_3(\bk)\equiv \sigma^\text{H}_0\,\bigg(1+\fract{F^1_{2}}{2}\bigg)\,,
\label{Hall conductivity BHZ final}
\eal
where $\sigma^\text{H}_0$ is the bare value, i.e., neglecting the vertex corrections, and 
$F^1_{2}$ is a weighted average over the Fermi surfaces. Therefore, in the model 
\eqn{Ham} we can explicitly verify that the corrections to the anomalous Hall conductivity 
only derive from the quasiparticle Fermi surface, as we earlier conjectured. We believe that this 
result has a more general validity. 
\\ 
For completeness, the Drude weight $D_{xx}=D_{yy}=D$ can be readily found through \eqn{conductivity longitudinal}, 
\beal
D &= D_{0}\,\bigg(1 + \fract{F^{1}}{2}\bigg)\,,
\label{Drude BHZ}
\eal
where
\beal
D_{0} &= -\fract{e^2}{2V}\,\sum_\bk\,\sum_{\ell=1}^2\,
\fract{\partial f\big(\ep_\ell(\bk)\big)}{\partial\ep_\ell(\bk)}\;
\bnabla_\bk\,\ep_\ell(\bk)\cdot\bnabla_\bk\,\ep_\ell(\bk)= \fract{e^2}{2V}\,\sum_\bk\,\sum_{\ell=1}^2\,
f\big(\ep_\ell(\bk)\big)\, \nabla^2_\bk\,\ep_\ell(\bk)\,,
\label{bare Drude}
\eal
is the bare value, and $F^{1}$ is again a weighted average over the Fermi surfaces of 
$F^1_{0}(\bk)$ and $F^1_{3}(\bk)$, see 
\eqn{current vertex BHZ}. In this case, $F^{1}/2$ is the correction 
to the optical mass with respect to the quasiparticle effective one, defined through the last equation in \eqn{bare Drude}. \\

\noindent
We emphasise that the 
Fermi liquid corrections to both anomalous Hall conductivity \eqn{Hall conductivity BHZ final} and Drude weight \eqn{Drude BHZ} stem from the fact that 
we are studying the response to a uniform electric field, which has to be evaluated in the $\omega$-limit. As a consequence, the dressed current vertex that enters the response functions is $\bd{J}^\omega$ instead of $\bd{J}^q$, the latter being the one obtainable from the Ward-Takahashi identity. In a Fermi liquid, 
$\bd{J}^\omega\not=\bd{J}^q$ because the quasiparticle Fermi surface induces 
a non-analyticity at $\omega=q=0$, which is the key to the microscopic derivation of Landau's Fermi liquid theory \cite{Nozieres&Luttinger-1}. It is tantalising to interpret the correction $\bd{J}^\omega-\bd{J}^q$ as due to an effective  
dipole moment carried by the quasiparticles, as proposed in \cite{CHEN2017345}, even though we do not have a clear physical argument in support.
\\
To conclude, we mention that, if instead of a fully spin-polarised BHZ model as in \eqn{Ham}, we 
considered the same model without explicitly breaking time-reversal symmetry, or models sharing similar topological ingredients, we could still discuss  topological properties as the non-quantised intrinsic component of the spin Hall conductivity, see, e.g., \cite{Nagaosa-Science2003,Nagaosa-PRB2004}.

\section{Conclusions}
\label{Conclusions}

Landau's theory of topological Fermi liquids predicts that the residual interactions among the quasiparticles, the $f$-parameters, yield corrections not only to conventional 
thermodynamic susceptibilities and longitudinal transport coefficients, like  
the Drude weight, but also to the intrinsic anomalous Hall conductivity. The latter is therefore not  
expressible solely in terms of the Berry phases acquired by the quasiparticle Bloch states adiabatically evolving on the Fermi surface, as one would intuitively argue \cite{Haldane-PRL2004}, 
and of the bare Chern number of occupied bands. Both contributions to the anomalous Hall conductivity 
are in fact renormalized by the interaction with the quasiparticles at the chemical potential, simply reflecting the well known difference between the static and dynamic limits of linear response functions in metals. This result, though not unexpected, has required a substantial extension of the 
original Nozi\'eres and Luttinger microscopic derivation of conventional Fermi liquids \cite{Nozieres&Luttinger-1}, which we present in Appendix \ref{Formal}.\\

\noindent
The corrections to the anomalous Hall conductivity that we uncover is the metallic counterpart of recent results \cite{Andrea-PRB2023,PhysRevLett.131.236601} showing that the Chern number of two-dimensional quantum anomalous Hall insulators not necessarily coincides with the topological 
invariant \cite{Ishikawa-1,Ishikawa-2,PhysRevLett.105.256803,Zhang-PRB2012} corresponding to the winding number $W(G)$, also denoted as $N_3(G)$, of the map $(\ep,\bk)\to \hat{G}(i\ep,\bk) \in 
\text{GL}(n,\mathbb{C})$, where 
$\hat{G}(i\ep,\bk)$ is the fully-interacting Green's function. 
Indeed, as demonstrated in \cite{Andrea-PRB2023}, this winding number is equivalent to that of 
the map $(\ep,\bk)\to \hat{G}_*(i\ep,\bk)$, where $\hat{G}_*(i\ep,\bk)$ 
is obtained by filtering out from $\hat{G}(i\ep,\bk)$ the quasiparticle residue, see Appendix \ref{Quasiparticle Green's function}, 
and thus coincides with the quasiparticle Green's function 
\eqn{self-consistency 2} in the doped insulator. Correspondingly, 
$W(G_*)$ reduces upon doping to the non-quantised $\sigma_0^\text{H}$ in units of 
$e^2/2\pi$, and the corrections predicted in \cite{Andrea-PRB2023,PhysRevLett.131.236601} to the Fermi liquid ones we have just derived. We remark that, since $\hat{G}_*(i\ep,\bk)$ in the doped insulator has singular eigenvalues at $\ep=0$ on the quasiparticle Fermi surface, $W(G_*)$ is, strictly speaking, not anymore 
a genuine winding number, though it remains perfectly defined. With this proviso, we hereafter refer to $W(G_*)$ still as the winding number. 
\\
\begin{figure}[t]
\centerline{\includegraphics[width=0.7\textwidth]{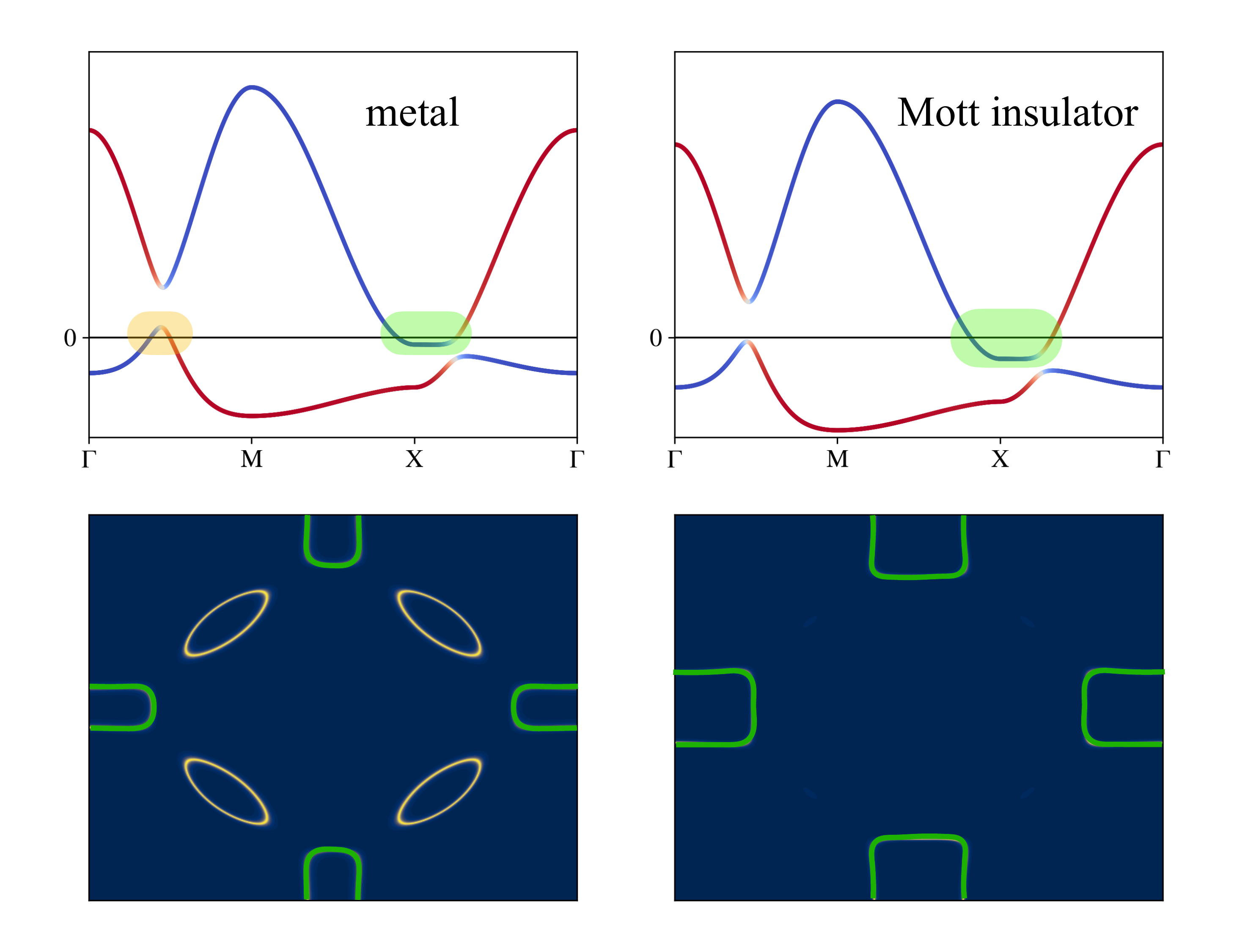}}
\caption{Left panels: hypothetical quasiparticle band structure in the metal phase (top) showing two distinct sets of quasiparticle Fermi pockets (bottom), one representing physical Fermi pockets, in yellow, which account for the hole doping, 
and the other, in green, being Luttinger pockets. Right panels: the same as 
left ones but on the Mott insulating side at $\delta=0$. Here, only the Luttinger pockets exist.}
\label{Fig2}
\end{figure}

\noindent
In the following, we show that, elaborating on the previous observation, 
one can draw interesting and rather general conclusions. 
We start emphasising that Refs.~\cite{Andrea-PRB2023} and \cite{PhysRevLett.131.236601} rationalise the puzzling evidence of Mott insulators with, 
presumably \cite{Note1}, vanishing Chern number and yet finite winding number due to the existence of 
in-gap topological bands of Green's function zeros \cite{Slagle-PRB2015,Giorgio-NatComm2023,Phillips-PRL2023,Qimiao-oscillations,Qimiao-zeros,Ivan-BHZ}. 
It is reasonable to conjecture that doping such Mott insulators with 
$\sigma^\text{H}=0$ but $W(G_*)\propto\sigma_0^\text{H}\not=0$ may lead, without any change of symmetry, to topological metals 
with finite $\sigma^\text{H}$.
For simplicity, we assume that such metal can be described as in 
Sect. \ref{A toy model calculation}, and use the results of that section to 
infer how it may continuously transform into the Mott insulator as the doping $\delta\to 0$.\\
The first and most obvious temptation is to assume that $\delta\to 0$ simply 
corresponds to a quasiparticle Fermi surface that shrinks into a point and disappears in the Mott insulator. 
In that case, since the bare $\sigma_0^\text{H}$ is expected to reach its quantised maximum magnitude when the valence band is full and the conduction one empty, the only possibility for $\sigma^\text{H}$ in 
\eqn{Hall conductivity BHZ final} to vanish is that $F^1_{2}\to -2$ for $\delta\to 0$. 
However, from the definition of $F^1_{2}(\bk)$ in \eqn{current vertex BHZ}, we would rather  
conclude that $F^1_{2}\to 0$ for $\delta\to 0$, unless $f^1_{2a}(\bk,\bkp)$, $a=0,3$, diverge sufficiently fast to compensate the vanishing Fermi volume. 
As we earlier discussed, such possibility has to be discarded, which implies 
that the assumption of a quasiparticle Fermi surface that 
disappears when $\delta\to 0$ is not consistent with $\sigma^\text{H}\to 0$, 
hence that a quasiparticle Fermi surface does survive till $\delta= 0$ 
to enforce $F^1_{2}\to -2$. Since quasiparticle Fermi surfaces, 
poles of $\text{det}\big(\hat{G}_*(0,\bk)\big)$,
comprise both physical Fermi and Luttinger surfaces \cite{mio-2,mio-Mott}, poles and zeros of 
$\text{det}\big(\hat{G}(0,\bk)\big)$, respectively, see also Appendix \ref{Quasiparticle Green's function}, 
we must conclude that the quasiparticle Fermi surface that exists at the 
transition into the Mott insulator is actually a Luttinger surface, which is allowed also in Mott insulators. 
This, in turn, implies that one of the in-gap bands of zeros of the Mott insulator right after the transition must cross the chemical potential and thus form a Luttinger surface, which is the smooth evolution of that one in the weakly doped metal phase.
In Fig.~\ref{Fig2} we show a hypothetical quasiparticle band structure 
that realises the above scenario, obtained by properly tuning the parameters 
in \eqn{Ham}. The metal phase, left panels, exhibits both Fermi pockets, in yellow, which account just for the hole doping \cite{mio-3}, and Luttinger ones, in green. Only the latter remains in the 
Mott insulator at $\delta=0$, right panels. \\
This physical scenario, which we have unveiled by quite general arguments, is 
fully consistent with the results of \cite{Andrea-PRB2023,PhysRevLett.131.236601}, since it 
predicts a Luttinger surface in the Mott insulator which is known to yield a violation of 
Luttinger's theorem \cite{Heath_2020,mio-3}. Such violation directly explains, by means of the Streda formula \cite{Streda_1982,Streda_1983}, why in the insulator
$\sigma^\text{H}$ can vanish despite $W(G)\not=0$. We also mention that recent 
numerical simulations of model interacting topological insulators \cite{Andrea-TKI,Ivan-BHZ}
find evidence of bands of Green's function zeros in the topological phase before the Mott 
transition, and which, should they cross the chemical potential and form a Luttinger surface, could    
provide a simple explanation \cite{Andrea-TKI} for the intriguing properties of SmB$_6$ and YbB$_{12}$ topological Kondo insulators. \\

\noindent 
If we take for granted the existence of a Luttinger surface in the topological metal at $\delta\ll 1$, we can draw a further conclusion. 
When $\delta\to 0$, the Drude weight \eqn{Drude BHZ} must vanish. However, $D_{0}$ in \eqn{bare Drude} is finite for $\delta\to 0$ because of the Luttinger surface, which implies that $D\to 0$ because $F^{1}\to -2$. In other words, while the quasiparticle effective mass,  
defined by the last equality in \eqn{bare Drude}, is smooth approaching the Mott phase, the optical mass diverges.  
Remarkably, this is precisely the scenario that might occur in a gapless quantum spin liquid \cite{mio-Mott}, see also \cite{RevModPhys.89.025003} and references therein, as well as the one that Grilli and Kotliar uncovered \cite{Marco&Gabi} in the $t-J$ model by a large-$N$ expansion around the saddle point within the slave-boson formalism. The analogy suggests that also in the weakly doped $t-J$ model there must be a Luttinger surface, and, possibly, also in other weakly doped Mott insulators, irrespective whether topology is involved. The advantage of the latter is that it allows straight reaching this conclusion because 
the winding number, a topological property of the Green's function, can well be finite in both metal and insulator.

\section*{Acknowledgements}
We are very grateful to Raffaele Resta for useful discussions and comments. 



\begin{appendix}
\section{Precise definition of the Fermi-liquid response functions}
\label{Appendix}

The correspondence between the low frequency, long wavelength and low temperature 
physical response functions and those obtained from the Hamiltonian 
\eqn{Hqp} treated by HF+RPA is in reality not straight. Indeed, that correspondence can be drawn only 
in the limits where one can make use of the Ward-Takahashi identities \cite{Nozieres&Luttinger-1}, therefore only in connection with densities associated to conserved quantities and their corresponding currents, and either in the $q$-limit or in the $\omega$-limit depending on the specific response function.\\
This point is particularly important in multiband models and for the current-current response function. 
Let us therefore begin by discussing the latter \cite{Nozieres&Luttinger-1,libro}. The Bethe-Salpeter equation for the reducible 
vertex in terms of the $f$-parameters is 
\beal
\Gamma_{\ell m,np}(\bk,\bkp;i\omega,\bq) &= f_{\ell m,np}(\bk,\bkp) \\
&\qquad + \fract{1}{V}\,\sum_\bp\, \sum_{q s}\, f_{\ell s, q p}(\bk,\bp) \;
R_{sq}(\bp;i\omega,\bq)\;
\Gamma_{q m,ns}(\bp,\bkp;i\omega,\bq)\,,
\label{Gamma}
\eal
where $\omega$ and $\bq$ are, respectively, the frequency and momentum transferred in the particle-hole channel, and the kernel 
\beal
R_{sq}(\bp;i\omega,\bq) &=T\sum_\ep\,\fract{1}{\;i\ep+i\omega-\ep_q(\bp+\bq)\;}
\;\fract{1}{\;i\ep-\ep_s(\bp)\;}\\
&=\fract{\;f\big(\ep_s(\bp)\big) - f\big(\ep_q(\bp+\bq)\big)\;}
{\; i\omega + \ep_s(\bp) - \ep_q(\bp+\bq)\;} \;.\label{R-App}
\eal
Equation~\eqn{Gamma} can be shortly written as 
\beal
\Gamma &= f + f\odot R\odot \Gamma = f + \Gamma\odot R\odot f\,,
\label{A1}
\eal
where the symbol $\odot$ denotes sum of internal indices and momenta. Hereafter, we make a series of 
formal and exact manipulations of the Bethe-Salpeter equation similar to those exploited by 
Nozi\'eres and Luttinger \cite{Nozieres&Luttinger-1} to derive Landau's Fermi liquid theory. 
We decided to show them explicitly since they might be not familiar to everybody. \\  
From \eqn{A1} we get
\beal
f &= \Gamma\odot\big(1+ R\odot \Gamma\big)^{-1}\,.
\label{A2}
\eal
We can take the $q$-limit of \eqn{A1}, sending first $\omega\to 0$ and then 
$\bq\to\bd{0}$, 
\beal
\Gamma^q &= f + \Gamma^q\odot R^q\odot f = f + f\odot R^q\odot\Gamma\,,
\label{A3}
\eal
and find 
\beal
f &= \big(1+\Gamma^q\odot R^q\big)^{-1}\odot\Gamma^q\,.
\label{A4}
\eal
Comparing \eqn{A2} with \eqn{A4} we obtain
\bealn
\Gamma\,\big(1+ R\odot \Gamma\big)^{-1} = \big(1+\Gamma^q\odot R^q\big)^{-1}\,\Gamma^q\,,
\eal
which implies 
\bealn
\big(1+\Gamma^q\odot R^q\big)\,\Gamma = \Gamma^q\,\big(1+ R\odot \Gamma\big)\,,
\eal
hence
\beal
\Gamma &= \Gamma^q + \Gamma^q\odot\big(R-R^q\big)\odot\Gamma 
= \Gamma^q + \Gamma\odot\big(R-R^q\big)\odot\Gamma^q\,.
\label{A5}
\eal
On the other hand, from \eqn{A5} it follows that 
\bealn
\Gamma^q &= \Gamma\odot\Big[1+\big(R-R^q\big)\odot\Gamma\Big]^{-1}\,,
\eal
and so, 
\beal
1 + R^q\odot\Gamma^q &= 1 + R^q\odot\Gamma\odot\Big[1+\big(R-R^q\big)\odot\Gamma\Big]^{-1}\\
&= \Big[1+\big(R-R^q\big)\odot\Gamma\Big]\odot\Big[1+\big(R-R^q\big)\odot\Gamma\Big]^{-1} \\
&\qquad +R^q\odot\Gamma\odot\Big[1+\big(R-R^q\big)\odot\Gamma\Big]^{-1}\\
&= \big(1+R\odot\Gamma\big)\odot\Big[1+\big(R-R^q\big)\odot\Gamma\Big]^{-1}\,.
\label{A6}
\eal
The dressed current vertex $\bd{J}$ is obtained from the bare one $\bd{J}_0$, 
which we actually do not know,  
through the Bethe-Salpeter equation 
\beal
\bd{J} = \bd{J}_0 +\bd{J}_0\odot R \odot \Gamma= \bd{J}_0 +\Gamma\odot R \odot \bd{J}_0 \,,
\label{BS-J}
\eal
whose $q$-limit is $\bd{J}^q = \bd{J}_0 +\bd{J}_0\odot R^q \odot \Gamma^q$. 
As before, we can use \eqn{BS-J} to solve for the unknown $\bd{J}_0$, 
\beal
\bd{J}_0 &= \bd{J}\odot\big(1+R \odot \Gamma\big)^{-1}\,,
\label{A7}
\eal
and, taking the $q$-limit as well as making use of \eqn{A6}, 
\beal
\bd{J}_0 &= \bd{J}^q\odot\big(1+R^q \odot \Gamma^q\big)^{-1}\\
&= \bd{J}^q\odot\Big[1+\big(R-R^q\big)\odot\Gamma\Big]
\odot\big(1+R\odot\Gamma\big)^{-1}\\
&= \big(1+\Gamma\odot R\big)^{-1}\odot\Big[1+\Gamma\odot\big(R-R^q\big)\Big]
\odot\bd{J}^q\,,
\label{A8}
\eal
which, compared with \eqn{A7}, leads to 
\beal
\bd{J} &= \bd{J}^q + \bd{J}^q\odot\big(R-R^q\big)\odot\Gamma
= \bd{J}^q + \Gamma\odot \big(R-R^q\big)\odot \bd{J}^q
\,.
\label{A9}
\eal
The current-current response function is defined, still in short notations, through 
\beal
\chi_{\bd{J}\bd{J}} &\equiv \Tr\Big(\bd{J}\odot R\odot\bd{J}_0\Big) 
= \Tr\Big(\bd{J}_0\odot R\odot\bd{J}\Big) \,,\label{A10}
\eal
whose $q$-limit is therefore $\chi^q = \Tr\Big(\bd{J}^q\odot R^q\odot\bd{J}_0\Big)$. 
We can thus write, making use of \eqn{BS-J} and \eqn{A9}, 
\beal
\chi_{\bd{J}\bd{J}} &= \chi^q + \Tr\Big(\bd{J}\odot R\odot\bd{J}_0\Big) 
-\Tr\Big(\bd{J}^q\odot R^q\odot\bd{J}_0\Big)\\
&= \chi^q + \Tr\Big(\bd{J}^q\odot \big(R-R^q\big)\odot\bd{J}_0\Big) + 
\Tr\Big(\bd{J}^q\odot \big(R-R^q\big)\odot\Gamma\odot R\odot\bd{J}_0\Big)\\
&= \chi^q + \Tr\Big(\bd{J}^q\odot \Delta\odot\bd{J}\Big)\\
&= \chi^q + \Tr\Big(\bd{J}^q\odot \Delta\odot\bd{J}^q\Big)+ \Tr\Big(\bd{J}^q\odot \Delta\odot\Gamma\odot \Delta\odot\bd{J}^q\Big)\,,
\label{current-current final}
\eal
where we have defined $\Delta = R-R^q$. We can now finally make use of symmetries and  of the Ward-Takahashi identities. We observe that gauge symmetry entails that $\chi^q$ must precisely cancel 
the diamagnetic term, which we also do not know, and the Ward-Takahashi identities that 
$\bd{J}^q(\bk) \equiv \bnabla_\bk\,\hat{H}_*(\bk)$ \cite{Nozieres&Luttinger-1}. 
Correspondingly, the component $\sigma_{ab}(i\omega,\bq)$ of the conductivity tensor reads   
\beal 
\sigma_{ab}&=  i\,e^2\,\lim_{\omega\to 0}\,\lim_{\bq\to\bd{0}}\,\fract{1}{i\omega}\;
\bigg\{\Tr\Big(J_a^q\odot \Delta \odot J_b^q\Big)
+ \Tr\Big(J_a^q\odot \Delta\odot\Gamma\odot\Delta\odot J_b^q\Big)\bigg\}\,,
\label{conductivity tensor}
\eal
thus involving the $\omega$-limit, first $\bq\to\bd{0}$ and then $\omega\to 0$. \\
It is worth noticing that in \eqn{current-current final} the kernel $R$ is substituted by $\Delta=R-R^q$. 
In other words, we have been obliged to manipulate RPA in order to build 
a correspondence with the physical response function and get rid of the unknown bare current vertex and diamagnetic term. 
From the explicit expression of $R$ in \eqn{R-App}, we find that 
\beal
\Delta_{\ell m}(\bk;\omega,\bq) &= 
R_{\ell m}(\bk;\omega,\bq) - R^q_{\ell m}(\bk) \\
&= 
\fract{\;f\big(\ep_m(\bk)\big)-f\big(\ep_\ell(\bk+\bq)\big)\;}{i\omega + \big(\ep_m(\bk)
-\ep_\ell(\bk+\bq)\big)}-\lim_{\bq\to 0}\,\fract{\;f\big(\ep_m(\bk)\big)-f\big(\ep_\ell(\bk+\bq)\big)\;}{\ep_m(\bk)
-\ep_\ell(\bk+\bq)}\;.
\label{A13}
\eal
Recalling that we are in any case interested in small $\bq$ and $\omega$, then 
\eqn{A13} for $\ell=m$ is at leading order, and defining $\bd{v}_\ell(\bk) = \bnabla_\bk\,\ep_\ell(\bk)$ 
the group velocity,  
\beal
\Delta_{\ell \ell}(\bk;\omega,\bq) \simeq 
-\fract{\partial f\big(\ep_\ell(\bk)\big)}{\partial  \ep_\ell(\bk)}\;
\fract{i\omega}{\;i\omega-\bd{v}_\ell(\bk)\cdot\bq\;}\;,
\label{R^q ell ell}
\eal
which is the standard result in the single-band case \cite{Nozieres&Luttinger-1}. 
On the contrary, for $\ell\not=m$ and assuming $\ep_\ell(\bk)\not=\ep_m(\bk)$, we readily find that 
\beal
\Delta_{\ell m}(\bk;\omega,\bq) \simeq -i\omega\; 
\fract{\;f\big(\ep_m(\bk)\big)-f\big(\ep_\ell(\bk)\big)\;}{\big(\ep_\ell(\bk)
-\ep_m(\bk)\big)^2}\;,
\label{R ell m}
\eal
which vanishes linearly in $\omega$. Since the $\omega$-limit 
appears in the conductivity, then 
\beal
\Delta^\omega_{\ell m}(\bk)&= \lim_{\omega\to 0}\,\lim_{\bq\to\bd{0}}\,\Delta_{\ell m}(\bk,\omega,\bq)
= -\delta_{m\ell}\;\fract{\partial f\big(\ep_\ell(\bk)\big)}{\partial\ep_\ell(\bk)}\;,\\
\dot{\Delta}^\omega_{\ell m}(\bk)&= \lim_{\omega\to 0}\,\lim_{\bq\to\bd{0}}\,\fract{\partial \Delta_{\ell m}(\bk,\omega,\bq)}{\partial i\omega}
= - \big(1-\delta_{m\ell}\big)\,
\fract{\;f\big(\ep_m(\bk)\big)-f\big(\ep_\ell(\bk)\big)\;}{\big(\ep_\ell(\bk)
-\ep_m(\bk)\big)^2}
\;.
\label{various limits}
\eal
In other words, the matrix $\Delta^\omega$ is diagonal while its 
derivative $\dot{\Delta}^\omega$ with respect to $\omega$ and calculated 
at $\omega=0$ is off-diagonal. This result is important in the calculation of the 
Hall conductivity. \\

\noindent
However, let us at first calculate the longitudinal conductivity, i.e., the variation of the electric current in the same direction, e.g., $x$, of a uniform and static electric field, divided by the field strength.   
From \eqn{conductivity tensor}, moving to the real frequency axis, 
$i\omega\to \omega+i0^+$ with small $\omega$ and setting $\bq=\bd{0}$, we 
find
\beal 
\sigma_{xx}(\omega) &=  i\,\fract{e^2}{\;\omega+i0^+\;}\,
\bigg\{\Tr\Big(J_x^q\odot \Delta^\omega \odot J_x^q\Big)
+ \Tr\Big(J_x^q\odot \Delta^\omega\odot\Gamma^\omega\odot\Delta^\omega\odot J_x^q\Big)\bigg\}\\
&=  i\,\fract{1}{\;\omega+i0^+\;}\;
e^2\,\Tr\Big(J_x^q\odot \Delta^\omega \odot J_x^\omega\Big)
\equiv i\,\fract{1}{\;\omega+i0^+\;}\;D_{xx}\,,
\label{conductivity longitudinal}
\eal
where $D_{xx}$ is the Drude weight and $\bd{J}^\omega$ the $\omega$-limit of 
the current vertex that, through \eqn{A9}, satisfies
\beal
\bd{J}^\omega &= \bd{J}^q + \bd{J}^q\odot\Delta^\omega\odot\Gamma^\omega
= \bd{J}^q + \Gamma^\omega\odot \Delta^\omega\odot \bd{J}^q
\,.
\label{Hall 4}
\eal
Let us now calculate the Hall conductivity through the 
matrix elements $\sigma_{ab}$ of \eqn{conductivity tensor}. For that we can simply adapt the results in \cite{Andrea-PRB2023}, which imply that 
the antisymmetric combination  
$(\sigma_{ab}-\sigma_{ba})/2$ defines the element $\sigma_{ab}^\text{H}$ of the 
anomalous Hall conductivity, which can be shown are simply  
\beal
\sigma_{ab}^\text{H} &= i\,e^2\,\lim_{\omega\to 0}\,\lim_{\bq\to\bd{0}}\,
\fract{\partial}{\partial i\omega}\,\bigg\{\Tr\Big(J_a^q\odot \Delta \odot J_b^q\Big)+ \Tr\Big(J_a^q\odot \Delta\odot\Gamma\odot\Delta\odot J_b^q\Big)\bigg\}\,.
\label{Hall 1}
\eal
Therefore, 
\beal
\sigma_{ab}^\text{H} &= i\,e^2\,\bigg\{ \Tr\Big(J_a^q\odot \dot{\Delta}^\omega \odot J_b^q\Big) + \Tr\Big(J_a^q\odot \dot{\Delta}^\omega\odot\Gamma^\omega\odot\Delta^\omega\odot J_b^q\Big)\\
&\qquad\qquad +\Tr\Big(J_a^q\odot \Delta^\omega\odot\Gamma^\omega\odot\dot{\Delta}^\omega\odot J_b^q\Big)
 +\Tr\Big(J_a^q\odot \Delta^\omega\odot\dot{\Gamma}^\omega\odot\Delta^\omega\odot J_b^q\Big)
\bigg\}\,,\label{Hall 2}
\eal
involving $\Delta^\omega$ and $\dot{\Delta}^\omega$ in \eqn{various limits}.
We note that the $\omega$-limit of the derivative of the Bethe-Salpeter equation \eqn{A1} is  
\bealn
\dot{\Gamma}^\omega &= f\odot \dot{\Delta}^\omega\odot \Gamma^\omega
+ f\odot R^\omega\odot \dot{\Gamma}^\omega
\,,
\eal
which leads to 
\bealn
\dot{\Gamma}^\omega &=\Gamma^\omega\,\dot{\Delta}^\omega\,\Gamma^\omega\,,
\eal
and, substituted into \eqn{Hall 2}, to the compact expression 
\beal
\sigma_{ab}^\text{H} &= i\,e^2\,\Tr\Big( J^\omega_{a}
\odot\dot{\Delta}^\omega
\odot J^\omega_{b}\Big)\,,\label{Hall 3}
\eal
which involves again the $\omega$-limit of the current vertex \eqn{Hall 4}.\\

\noindent
Let us finally consider the charge density-density response function. We can repeat step-by-step the 
above manipulations but now using the $\omega$-limit of the vertex. The final result is that the charge density-density response function reads 
\beal 
\chi_{\rho\rho}  
&= \chi_{\rho\rho}^\omega + \Tr\Big(\rho^\omega\odot \big(R-R^\omega\big)\odot\rho^\omega\Big)
+ \Tr\Big(\rho^\omega\odot \big(R-R^\omega\big)\odot\Gamma\odot\big(R-R^\omega\big)\odot\rho^\omega\Big)\,,
\label{A14}
\eal
where $\rho^\omega$ is the $\omega$-limit of the charge density vertex,  
\beal
\chi_{\rho\rho}^\omega &=  \Tr\Big(\rho^\omega\odot R^\omega\odot \rho_0\Big)
\,,
\label{A15}
\eal
the $\omega$-limit of the response function, and $\rho_0$ the bare vertex. Charge conservation implies that $\chi_{\rho\rho}^\omega=0$, while the Ward-Takahashi identities that $\rho^\omega$ is the 
identity matrix in orbital space \cite{Nozieres&Luttinger-1}. 
Since $R^q-R^\omega=-\Delta^\omega$, see \eqn{various limits}, 
the charge compressibility is readily obtained as 
\beal
\kappa &= -\chi^q_{\rho\rho} = \Tr\Big(\Delta^\omega\Big)
- \Tr\Big(\Delta^\omega\odot\Gamma^q\odot\Delta^\omega\Big)\\
&= -\fract{1}{V}\,\sum_{\bk\ell}\, \fract{\partial f\big(\ep_\ell(\bk)\big)}{\partial  \ep_\ell(\bk)}
-\fract{1}{V^2}\,\sum_{\bk\bkp\ell\ell'}\, \fract{\partial f\big(\ep_\ell(\bk)\big)}{\partial  \ep_\ell(\bk)}\;\Gamma^q_{\ell\ell',\ell'\ell}(\bk,\bkp)\;
\fract{\partial f\big(\ep_{\ell'}(\bkp)\big)}{\partial  \ep_{\ell'}(\bkp)}\;,
\label{compressibility}
\eal
which is the conventional result of Fermi liquid theory showing that the compressibility is 
not just the quasiparticle density-of-states at the chemical potential, but 
acquires a correction from the 
quasiparticle interaction.

\section{Quasiparticle Green's function}
\label{Quasiparticle Green's function}

In this Appendix we show how to rigorously define the quasiparticle Green's function $\hat{G}_*(i\ep,\bk)$ through the fully interacting thermal one $\hat{G}(i\ep,\bk)$ of the physical electrons, having in mind a multiband system in which both Green's functions are matrices. This will give us the opportunity to clarify some 
results mentioned in the main text.
\\
The physical Green's function $\hat{G}(i\ep,\bk)$ satisfies the Dyson equation
\bealn
\hat{G}(i\ep,\bk)^{-1} &= i\ep -\hat{H}_0(\bk) - \hat{\Sigma}(i\ep,\bk)\,,
\eal
where $\hat{H}_0(\bk)$ is the non-interacting Hamiltonian, not to be confused with the quasiparticle one in \eqn{Hqp}, and the self-energy $\hat{\Sigma}(i\ep,\bk)$ 
accounts for all interaction effects. 
We recall that $\hat{\Sigma}(i\ep,\bk)^\dagger = \hat{\Sigma}(-i\ep,\bk)$, 
which implies that  
\bealn
\hat{\Sigma}_1(i\ep,\bk) &= \fract{1}{2}\,\Big(\hat{\Sigma}(i\ep,\bk) + 
\hat{\Sigma}(i\ep,\bk)^\dagger\Big) = \fract{1}{2}\,\Big(\hat{\Sigma}(i\ep,\bk) + 
\hat{\Sigma}(-i\ep,\bk)\Big)\,,
\eal
is hermitean and even in $\ep$, while
\bealn
\hat{\Sigma}_2(i\ep,\bk) &= \fract{1}{2i}\,\Big(\hat{\Sigma}(i\ep,\bk) -
\hat{\Sigma}(i\ep,\bk)^\dagger\Big) = \fract{1}{2i}\,\Big(\hat{\Sigma}(i\ep,\bk) -
\hat{\Sigma}(-i\ep,\bk)\Big)\,,
\eal
is still hermitean but odd in $\ep$. In addition, its eigenvalues are negative for positive $\ep$,  
positive for negative $\ep$, and vanish when $\ep$ is strictly zero.\\
One defines \cite{mio-Mott} a semi-positive definite matrix 
\beal
\hat{Z}(\ep,\bk) &= \Bigg(1 - \fract{\;\hat{\Sigma}_2(i\ep,\bk)\;}{\ep}\Bigg)^{-1} 
= \hat{A}(\ep,\bk)^\dagger\, \hat{A}(\ep,\bk)\,,\label{Z(epsilon)}
\eal
with eigenvalues $\in [0,1]$, which plays the role of the quasiparticle residue, and 
the hermitian frequency-dependent Hamiltonian
\beal
\hat{H}_*(\ep,\bk) &=  \hat{H}_*(-\ep,\bk)=\hat{A}(\ep,\bk)\,\Big(\hat{H}_0(\bk) + \hat{\Sigma}_1(i\ep,\bk)\Big)\,
\hat{A}(\ep,\bk)^\dagger\,,\label{H*(epsilon)}
\eal
through which 
\beal
\hat{G}(i\ep,\bk) &= \hat{A}(\ep,\bk)^\dagger\;\fract{1}{\;i\ep-\hat{H}_*(\ep,\bk)\;}\;
\hat{A}(\ep,\bk) \,.\label{G(epsilon)}
\eal
The quasiparticle Green's function \eqn{self-consistency 2} is simply obtained through the 
low-frequency limit of $\big(i\ep-\hat{H}_*(\ep,\bk)\big)^{-1}$ in \eqn{G(epsilon)}, which, provided  
$\hat{H}_*(\ep,\bk) = \hat{H}_*(-\ep,\bk)$ has a regular Taylor expansion in $\ep$, i.e.,  
$\hat{H}_*(\ep,\bk)\simeq \hat{H}_*(0,\bk) + O\left(\ep^2\right)$, 
reads, at leading order, 
\beal
\hat{G}_*(i\ep,\bk) &\simeq \fract{1}{\;i\ep-\hat{H}_*(0,\bk)\;}
\equiv \fract{1}{\;i\ep-\hat{H}_*(\bk)\;}\;.
\label{App: G_*}
\eal
The \textit{quasiparticle Fermi surface} corresponds to the manifold $\bk=\bk_{*F}$ in momentum space where 
\beal
\text{det}\Big(\hat{H}_*(\bk_{*F})\Big) = \text{det}\Big(\hat{Z}(0,\bk_{*F})\Big)\, 
\text{det}\Big(\hat{H}_0(\bk_{*F}) + \hat{\Sigma}_1(0,\bk_{*F})\Big) = 0\,,
\label{App: definition Fermi surface}
\eal
which includes the roots of both terms on the right hand side, and where $\hat{Z}(0,\bk)$ 
must be evaluated through the limit $\ep\to 0$ of \eqn{Z(epsilon)}. If $\text{det}\big(\hat{Z}(0,\bk)\big)$ is finite, we observe that, since 
\bealn
\hat{G}(0,\bk) &= -\fract{1}{\;\hat{H}_0(\bk) + \hat{\Sigma}_1(0,\bk)\;}\,,
\eal
the roots of $\text{det}\big(\hat{H}_0(\bk) + \hat{\Sigma}_1(0,\bk)\big)$ are 
actually the poles of $\text{det}\big(\hat{G}(0,\bk)\big)$ and define the physical 
Fermi surface, $\bk=\bk_F$. On the contrary, one realises through \eqn{G(epsilon)} that the roots of 
$\text{det}\big(\hat{Z}(0,\bk)\big) = \big|\text{det}\big(\hat{A}(0,\bk)\big)\big|^2$ 
correspond to those of $\text{det}\big(\hat{G}(0,\bk)\big)$, which thus define the physical Luttinger 
surface $\bk=\bk_L$. More precisely, $\text{det}\big(\hat{\Sigma}_1(0,\bk)\big)$ has a simple pole on 
the Luttinger surface, thus $\text{det}\big(\hat{G}(0,\bk_L)\big)=0$, while 
$\text{det}\big(\hat{Z}(0,\bk)\big)$ a second order root, so that \eqn{App: definition Fermi surface} 
is indeed verified for $\bk_{*F}=\bk_L$. It follows that the \textit{quasiparticle Fermi surface} 
$\bk_{*F}= \bk_F\cup \bk_L$ comprises both physical Fermi and Luttinger surfaces, as we mentioned 
in the main text. \\

\noindent
In reality, \eqn{G(epsilon)} is an exact factorisation of the physical electron Green's function that remains valid also in insulators lacking 
Fermi and Luttinger surfaces, and which was exploited in \cite{Andrea-PRB2023} to explicitly calculate 
the winding number $W(G)$ in two dimensions. Indeed, since $W(M\,N)=W(M)+W(N)$ and $W(M)=0$ if 
$M=M^\dagger$, then 
\bealn
W(G) &= W(A^\dagger\,G_*\,A) = W(G_*) + W(A^\dagger\,A) = W(G_*)\,.
\eal
One can rigorously prove \cite{Andrea-PRB2023} that $W(G_*)$ reduces to the well-known TKNN formula \cite{TKNN} 
calculated with the eigenstates of $\hat{H}_*(\bk)$ in \eqn{App: G_*}. The derivation remains simply true even if there is a quasiparticle Fermi surface, a result we used in the Sect.~\ref{Conclusions}.

\section{Formal derivation of Landau's Fermi liquid theory}
\label{Formal}
Assuming the case in which a quasiparticle Fermi surface exists, Landau's Fermi liquid theory 
can be microscopically derived \cite{Nozieres&Luttinger-1,libro} under the sole condition   
that $\hat{Z}(\ep,\bk)$ and 
$\hat{H}_*(\ep,\bk)$ are analytic matrix-valued functions of $\ep$ in the vicinity of the origin $\ep=0$ 
and of $\bk$ close to the quasiparticle Fermi surface $\bk_{*F}$. This condition is verified not only 
near a physical Fermi surface, $\bk\simeq \bk_F$, but also near a Luttinger surface \cite{mio-2,mio-Mott}, $\bk\simeq \bk_L$, despite the singular self-energy. Moreover, we emphasise that the Nozi\`eres and Luttinger derivation \cite{Nozieres&Luttinger-1} is fully non-perturbative, and, therefore, perfectly valid also when a Luttinger surface is present, which does entail  
a breakdown of perturbation theory. \\


\noindent
Since most of the technical details have been already presented in Appendix \ref{Appendix}, 
we end by briefly sketching how Nozi\`eres and Luttinger derivation \cite{Nozieres&Luttinger-1} 
works to appreciate its elegance and non-perturbative character. Moreover, 
we attempt a generalisation of the theory to a multi-band Fermi liquid to 
assess under which conditions the results in Appendix \ref{Appendix} do reproduce 
the physical thermodynamic susceptibilities and transport coefficients.\\
Let us therefore consider again 
the Bethe-Salpeter equations \eqn{A1} and \eqn{BS-J} for the scattering four-leg vertex $\Gamma$  
and the current, $\bd{J}$, or the density, $\rho$, vertices, now, however, 
written for the physical electrons. For convenience, we hereafter focus 
on the current-current response function, since the extension to the 
density-density one is straightforward. 
The Bethe-Salpeter equations involve a kernel $R$, which is a tensor that depends 
on four indices in the chosen basis of single-particle wavefunctions. 
Specifically, 
\beal
&R_{\alpha\beta,\gamma\delta}(i\ep,\bk;i\omega,\bq) = 
G_{\alpha\beta}(i\ep+i\omega,\bk+\bq)\,G_{\gamma\delta}(i\ep,\bk)\\
&\quad = \bigg(\fract{1}{i\ep+i\omega-H_0(\bk+\bq)-\Sigma(i\ep+i\omega,\bk+\bq)}\bigg)
_{\alpha\beta}\,
\bigg(\fract{1}{i\ep-H_0(\bk)-\Sigma(i\ep,\bk)}\bigg)_{\gamma\delta}\,.
\label{explain:R}
\eal 
We now use the exact representation of $\hat{G}(i\ep,\bk)$ in \eqn{G(epsilon)} 
and define the quasiparticle vertices $\Gamma_*$ and $\bd{J}_*$ by contracting 
each incoming external leg of the physical vertices 
with $\hat{A}(\ep,\bk)$ and each outgoing one with 
$\hat{A}(\ep,\bk)^\dagger$. The result is that in the Bethe-Salpeter equations \eqn{A1} and \eqn{BS-J} all vertices are replaced by the quasiparticle ones, 
and the kernel \eqn{explain:R} by
\beal
&R_{*\,\alpha\beta,\gamma\delta}(i\ep,\bk;i\omega,\bq) = 
\mathcal{G}_{\alpha\beta}(i\ep+i\omega,\bk+\bq)\,\mathcal{G}_{\gamma\delta}(i\ep,\bk)\\
&\qquad = \bigg(\fract{1}{i\ep+i\omega-\hat{H}_*(\ep+\omega,\bk+\bq)}\bigg)
_{\alpha\beta}\,
\bigg(\fract{1}{i\ep-\hat{H}_*(\ep,\bk)}\bigg)_{\gamma\delta}\,.
\label{explain:R*}
\eal
Without making any assumption whatsoever, one can formally manipulate the new 
Bethe-Salpeter equations, as discussed in Appendix \ref{Appendix}, so as to express them in terms 
of the vertices $\Gamma_*^q$ and $\bd{J}_*^q$ calculated in the static $q$-limit, 
where the latter is fully determined by the Ward-Takahashi identities \cite{Nozieres&Luttinger-1}.  
The outcome of such exact manipulation is that the kernel $R_*$ in \eqn{explain:R*} is replaced by $R_*-R_*^q$, which,   
following \cite{Nozieres&Luttinger-1}, we would like to represent in the sense of a distribution $\Delta_*$ in the Matsubara frequency $\ep$. 
Let us discuss separately the derivation of $\Delta_*$ and of its derivative 
$\partial_{i\omega}\,\Delta_*$ at $\omega=0$, starting from the former.

\subsection{Kernel for the longitudinal conductivity}
We note that $\hat{\mathcal{G}}(i\ep,\bk)$ is continuous for any $\ep\not=0$, 
while at $\ep=0$ its anti-hermitian part changes sign discontinuously. It 
follows that $R_*$ in \eqn{explain:R*} has two discontinuities at $\ep=0$ 
and $\ep=-\omega$, whereas $R_*^q$ a single one at $\ep=0$. Since we are 
interested in small $\omega$ and $\bq$, actually in their $\omega$-limit for 
the calculation of transport coefficients, we realise that the only region 
in $\ep$ where $R_*$ differs from $R_*^q$ is when $-\omega<\ep <0$, assuming for 
simplicity $\omega>0$. Since $\omega$ is small and $\hat{H}_*(\ep,\bk)$ 
is even in $\ep$, in this region we can approximate      
\beal
\hat{\mathcal{G}}(i\ep,\bk) &= 
\fract{1}{\;i\ep-\hat{H}_*(\ep,\bk)\;}\simeq \fract{1}{\;i\ep-\hat{H}_*(0,\bk)\;}
= \fract{1}{\;i\ep-\hat{H}_*(\bk)\;} = \hat{G}_*(i\ep,\bk)\;,
\label{explain:G*}
\eal
provided $\hat{H}_*(\ep,\bk)$ is analytic around $\ep=0$. It follows that, if 
$F(i\ep)$ is a test function smooth around $\ep=0$ and we formally write 
$R_* = \hat{\mathcal{G}}(i\ep+i\omega,\bk+\bq)\otimes \hat{\mathcal{G}}(i\ep,\bk)$ then 
\bealn
&T\sum_\ep\,\big(R_*-R_*^q\big)\,F(i\ep) \xrightarrow[\omega\text{-limit}]{}
T\sum_{-\omega<\ep<0}\,\hat{\mathcal{G}}(i\ep+i\omega,\bk)\otimes 
\hat{\mathcal{G}}(i\ep,\bk)\,
F(i\ep)\\
&\qquad \simeq F(0)\,T\sum_{-\omega<\ep<0}\,\hat{G}_*(i\ep+i\omega,\bk)\otimes \hat{G}_*(i\ep,\bk)\,.
\eal
Now, we can project onto the basis that diagonalises $\hat{H}_*(\bk)$, and, specifically, on the element $(\ell,m)$, so that 
\bealn
&F(0)\,T\sum_{-\omega<\ep<0}\,\hat{G}_*(i\ep+i\omega,\bk)\otimes \hat{G}_*(i\ep,\bk) \to F(0)\,T\sum_{-\omega<\ep<0}\,\fract{1}{i\ep+i\omega-\ep_\ell(\bk)}\;
\fract{1}{i\ep-\ep_m(\bk)}\\
&\qquad = F(0)\;\fract{1}{i\omega-\ep_\ell(\bk)+\ep_m(\bk)}\;
T\!\sum_{-\omega<\ep<0}\,\bigg(\fract{1}{i\ep-\ep_m(\bk)}-
\fract{1}{i\ep+i\omega-\ep_\ell(\bk)}\bigg)\,.
\eal 
Since the sum is proportional to $\omega$, a non-zero result for $\omega\to 0$ 
requires $\ell=m$, excluding the possibility of crossing bands. In this case
\bealn
F(0)\;\fract{1}{i\omega}\;
T\sum_{-\omega<\ep<0}\,\bigg(\fract{1}{i\ep-\ep_\ell(\bk)}-
\fract{1}{i\ep+i\omega-\ep_\ell(\bk)}\bigg) 
\xrightarrow[\omega\to 0]{} - F(0)\,\fract{\partial f\big(\ep_\ell(\bk)\big)}
{\partial \ep_\ell(\bk)}\,,
\eal
where $f(x)$ is the Fermi distribution function, which derives from 
\bealn
&T\sum_{-\omega<\ep<0}\,\bigg(\fract{1}{i\ep-x}-\fract{1}{i\ep+i\omega-x}\bigg)
=T\sum_{0<\ep<\omega}\,\bigg(\fract{1}{-i\ep-x}-\fract{1}{i\ep-x}\bigg)\\
&\qquad = T\sum_{\ep>0}\,\bigg(\fract{1}{i\ep+i\omega-x}-\fract{1}{i\ep-x}
+\fract{1}{-i\ep-x} -\fract{1}{-i\ep-i\omega-x}\bigg)\\
&\qquad \simeq -i\omega\,T\sum_{\ep>0}\,\bigg(
\fract{1}{(i\ep-x)^2} + \fract{1}{(-i\ep-x)^2}\bigg)
=-i\omega\,T\sum_\ep\,\fract{1}{(i\ep-x)^2} = -i\omega\,\fract{\partial f(x)}{\partial x}\;.
\eal
In conclusion, the distribution $\Delta_*$ in the $\omega$-limit is simply
\beal
\Delta^\omega_{*\,\ell m}(i\ep,\bk) = -\delta_{\ell m}\;\fract{\delta_{\ep 0}}{T}\;\fract{\partial f\big(\ep_\ell(\bk)\big)}
{\partial \ep_\ell(\bk)}\,,
\label{explain:Delta*-omega}
\eal
which coincides with \eqn{various limits} if we identify the eigenvalues 
$\ep_\ell(\bk)$ with the Hartree-Fock ones of the Hamiltonian \eqn{Hqp}. \\

\noindent
We recall that $\,\,\hat{\!\!\bd{J}}_*^q(\ep,\bk) = \bd{\nabla}_\bk\,\hat{H}_*(\ep,\bk)$ is hermitian and even in $\ep$, and, because of \eqn{explain:Delta*-omega}, 
\beal
\,\,\hat{\!\!\bd{J}}_*^\omega(\ep,\bk) &= 
\bd{\nabla}_\bk\,\hat{H}_*(\ep,\bk) +\fract{1}{V}\,\sum_\bkp\,
\hat{\Gamma}_*^\omega(i\ep,\bk;0,\bkp)\odot\hat{\Delta}^\omega(\bk')
\odot\bd{\nabla}_\bkp\,\hat{H}_*(\bkp)\\
&= \bd{\nabla}_\bk\,\hat{H}_*(\ep,\bk) + \delta\,\,\hat{\!\!\bd{J}}_*^\omega(\ep,\bk)\,,\\
\hat{\Gamma}_*^\omega(i\ep,\bk;0,\bkp) &= \hat{\Gamma}_*^q(i\ep,\bk;0,\bkp)
+ \fract{1}{V}\,\sum_\bp\,\hat{\Gamma}_*^q(i\ep,\bk;0,\bp) 
\odot\hat{\Delta}^\omega(\bp)
\odot\hat{\Gamma}_*^\omega(0,\bp;0,\bkp)\,,
\label{explain: current w-limit}
\eal
where $\hat{\Gamma}_*^\omega(0\bk,0\bkp)$ and $\hat{\Gamma}_*^q(0\bk,0\bkp)$ must be identified with the Landau $f$ and $A$ parameters, respectively. Therefore, 
$\hat{\Gamma}_*^{\omega/q}(\ep\bk,0\bkp)$ must be continuous across $\ep=0$, and 
thus 
\beal
\,\,\hat{\!\!\bd{J}}_*^\omega(\ep\to 0^+,\bk) &=
\,\,\hat{\!\!\bd{J}}_*^\omega(\ep\to 0^-,\bk) = \,\,\hat{\!\!\bd{J}}_*^\omega(\bk)
= \,\,\hat{\!\!\bd{J}}_*^\omega(\bk)^\dagger\,.
\label{explain: current w-limit ep=0}
\eal
We can elaborate on further. We note that $\,\,\hat{\!\!\bd{J}}_*^q(\ep,\bk)$ and $\hat{\Gamma}_*^q(i\ep,\bk;i\ep',\bkp)$
are related to each other by a Bethe-Salpeter equation. Since the outcome of this equation is $\,\,\hat{\!\!\bd{J}}_*^q(\ep,\bk)$, which is even in $\ep$, 
it is plausible, as can be indeed verified in the lowest order skeleton expansion, 
that $\hat{\Gamma}_*^q(i\ep,\bk;0,\bkp)$, and thus, through \eqn{explain: current w-limit}, $\hat{\Gamma}_*^\omega(i\ep,\bk;0,\bkp)$ are also even in $\ep$. 
We thus conclude that $\,\,\hat{\!\!\bd{J}}_*^\omega(\ep,\bk)=\,\,\hat{\!\!\bd{J}}_*^\omega(-\ep,\bk)
= \,\,\hat{\!\!\bd{J}}_*^\omega(\ep,\bk)^\dagger$ is hermitian. 
More explicitly, we find, through 
\eqn{explain: current w-limit} and \eqn{explain:Delta*-omega}, that 
\beal
&\Gamma_{*\,\ell n,n m}^\omega(i\ep,\bk;0,{\bkp}) = \Gamma_{*\,\ell n,n m}^q(i\ep,\bk;0,{\bkp})\\
&\qquad\qquad\qquad\qquad - \fract{1}{V}\,\sum_{r\bp}\,\Gamma_{*\,\ell r,r m}^q(i\ep,\bk;0,\bp) 
\fract{\partial f\big(\ep_r(\bp)\big)}{\partial \ep_r(\bp)}\;
\Gamma_{*\,r n,n r}^\omega(0,\bp;0,{\bkp})\\
&\qquad\qquad\qquad=\Gamma_{*\,\ell n,n m}^q(i\ep,\bk;0,{\bkp})- \fract{1}{V}\sum_{r\bp}\,
\Gamma_{*\,\ell r,r m}^q(i\ep,\bk;0,\bp) 
\fract{\partial f\big(\ep_r(\bp)\big)}{\partial \ep_r(\bp)}\,
f_{r\bp,n\bkp}\,,
\label{explain: explicit Gamma w-limit}
\eal
and thus 
\beal
&\delta \bd{J}_{*\,\ell m}^\omega(\ep,\bk) = 
-\fract{1}{V}\,\sum_{\bkp,n}\,
\Gamma^q_{*\,\ell n,n m}(i\ep,\bk;0,{\bkp})\;
\fract{\partial f\big(\ep_n(\bkp)\big)}{\partial \bkp}\\
&\qquad\qquad + \fract{1}{V^2}\,\sum_{\bkp \bp}\,\sum_{n r}\,
\Gamma_{*\,\ell r,r m}^q(i\ep,\bk;0,\bp) \,
\fract{\partial f\big(\ep_r(\bp)\big)}{\partial \ep_r(\bp)}\,\fract{\partial f\big(\ep_n(\bkp)\big)}{\partial \bkp}\,f_{r\bp,n\bkp}\,.
\label{explain: explicit delta J omega}
\eal

\subsection{Hall conductivity}

Since the distribution $\partial_{i\omega}\,\Delta_*\equiv \partial_{i\omega}\,R_*$ is needed to 
obtain the expression of the Hall conductivity \eqn{Hall 3}, we derive it by 
explicitly calculating the latter quantity closely following \cite{Andrea-PRB2023}.\\
For convenience, we here define 
\bealn
R_* &= \hat{\mathcal{G}}(i\ep+i\omega,\bk+\bq)\otimes 
\hat{\mathcal{G}}(i\ep-i\omega,\bk)\,,
\eal
so that 
\beal
\partial_{i\omega}\,\Delta_* &= \fract{\partial R_*}{\partial 2i\omega}\,
\xrightarrow[\omega\text{-limit}]{} \,-\fract{i}{2}\,\Big(
\partial_{\ep}\,\hat{\mathcal{G}}(i\ep,\bk)\otimes 
\hat{\mathcal{G}}(i\ep,\bk) - \hat{\mathcal{G}}(i\ep,\bk)\otimes 
\partial_{\ep}\,\hat{\mathcal{G}}(i\ep,\bk)\Big)\,.
\label{explain: def derivative Delta}
\eal
Therefore, dropping for simplicity the asterisks in the current vertices, the 1-2  component of the Hall conductivity \eqn{Hall 3} can be formally written as 
\beal
\sigma^\text{H}_{12} &= \fract{e^2}{2V}\,\sum_\bk\,T\sum_\ep\,\ep_{\mu\nu}\,
\Tr\Big(\hat{J}_\mu^\omega(\ep,\bk)\; \partial_{\ep}\,\hat{\mathcal{G}}(i\ep,\bk)\;
\hat{J}_\nu^\omega(\ep,\bk)\;\hat{\mathcal{G}}(i\ep,\bk)\Big)\\
&= -\fract{e^2}{2V}\,\sum_\bk\,T\sum_\ep\,\ep_{\mu\nu}\,
\Tr\Big(\hat{J}_\mu^\omega(\ep,\bk)\;\hat{\mathcal{G}}(i\ep,\bk)\; \partial_{\ep}\,\hat{\mathcal{G}}(i\ep,\bk)^{-1}\;\hat{\mathcal{G}}(i\ep,\bk)
\hat{J}_\nu^\omega(\ep,\bk)\;\hat{\mathcal{G}}(i\ep,\bk)\Big)\\
&= -\fract{e^2}{2V}\,\sum_\bk\,T\sum_\ep\,\ep_{\mu\nu}\,\sum_{\ell m n}\,
\mathcal{G}_\ell(i\ep,\bk)\,\mathcal{G}_m(i\ep,\bk)\,\mathcal{G}_n(i\ep,\bk)\\
&\qquad\qquad\qquad\qquad\qquad\qquad\qquad
 J^\omega_{\mu,\ell m}(\ep,\bk)\;\Big(\partial_\ep\,\hat{\mathcal{G}}(i\ep,\bk)^{-1}\Big)_{m n}\;
J^\omega_{\nu,n \ell}(\ep,\bk)\,,
\label{explain: step zero}
\eal
where the antisymmetric tensor $\ep_{\mu\nu}$ equals 1 if $\mu=1$ and $\nu=2$, and we use the basis that diagonalizes $\hat{H}_*(\ep,\bk)$, i.e., 
$\hat{H}_*(\ep,\bk)\ket{\!u_\ell(\ep,\bk)} = \ep_\ell(\ep,\bk)\ket{\!u_\ell(\ep,\bk)}$. One can readily verify that, because $\hat{\mathcal{G}}(i\ep,\bk)^\dagger
=\hat{\mathcal{G}}(-i\ep,\bk)$ and the current vertices are hermitian, 
\eqn{explain: step zero} is correctly real.\\
Since the last term in \eqn{explain: step zero} with all 
indices equal, $l=m=n$, vanishes, we can write
\beal
\sigma^\text{H}_{12} &=\fract{e^2}{2V}\,\sum_\bk\,T\sum_\ep\,\ep_{\mu\nu}\,\sum_{\ell \not= m}\,
\partial_\ep\,\mathcal{G}_m(i\ep,\bk)\,\mathcal{G}_\ell(i\ep,\bk)\;J^\omega_{\mu,\ell m}(\ep,\bk)\;
J^\omega_{\nu,m \ell}(\ep,\bk)\\
&\quad +\fract{e^2}{2V}\,\sum_\bk\,T\sum_\ep\ep_{\mu\nu}\sum_{\ell, m \not=n}
\mathcal{G}_\ell(i\ep,\bk)\,\mathcal{G}_m(i\ep,\bk)\,\mathcal{G}_n(i\ep,\bk)\,J^\omega_{\mu,\ell m}(\ep,\bk)\,F^\ep_{mn}(\ep,\bk)\,J^\omega_{\nu,n \ell}(\ep,\bk)\\
&= \sigma^\text{1H}_{12} + \sigma^\text{2H}_{12}\,,
\label{explain: sigma Hall 0}
\eal
where we define   
\bealn
F^\ep_{\ell m}(\ep,\bk) &= 
-\bra{u_\ell(\ep,\bk)} \partial_\ep\hat{\mathcal{G}}^{-1}(\ep,\bk)\ket{u_m(\ep,\bk)}\\
&=
\bra{u_\ell(\ep,\bk)} \partial_\ep\hat{H}_*(\ep,\bk)\ket{u_m(\ep,\bk)}\,,&
\ell&\not= m\,,\\
J^\omega_{\mu,\ell m}(\ep,\bk) &= \bra{u_\ell(\ep,\bk)} \hat{J}^\omega_\mu(\ep,\bk)\ket{u_m(\ep,\bk)}\,.
\eal
We note that, if we change the Hamiltonian $\hat{H}_*(\ep,\bk)$ into
\beal
\hat{H}_*(\ep,\bk;\bd{\kappa}) &= 
\hat{H}_*(\ep,\bk)  
+\bd{\kappa}\cdot\delta\,\,\hat{\!\!\bd{J}}^\omega(\ep,\bk)\,,
\label{explain: new Hamiltonian}
\eal
then, for $\ell\not=m$ and assuming the parallel-transport gauge, 
\bealn
F^\ep_{\ell m}(\ep,\bk) &= \big(\ep_{\ell}(\ep,\bk)-\ep_{m}(\ep,\bk)\big)\,
\bra{\partial_\ep u_\ell(\ep,\bk)}u_m(\ep,\bk)\rangle\\
&= \big(\ep_{m}(\ep,\bk)-\ep_{\ell}(\ep,\bk)\big)\,
\langle u_\ell(\ep,\bk)\ket{\partial_\ep u_m(\ep,\bk)}\,,\\
J^\omega_{\mu,\ell m}(\ep,\bk) &= \big(\ep_{\ell}(\ep,\bk)-\ep_{m}(\ep,\bk)\big)\,
\bra{\partial_\mu u_\ell(\ep,\bk)}u_m(\ep,\bk)\rangle\\
&= \big(\ep_{m}(\ep,\bk)-\ep_{\ell}(\ep,\bk)\big)\,
\langle u_\ell(\ep,\bk)\ket{\partial_\mu u_m(\ep,\bk)}\,,
\eal
with $\partial_\mu= \partial_{k_\mu} + \partial_{\kappa_\mu}$, and, not to weigh 
too much the notations, we do not explicitly indicate the $\bkappa$-dependence 
since in the final formulas the derivatives are evaluated at $\bkappa=\bd{0}$. \\
Coming back to the Hall conductivity, one easily realises that the term in \eqn{explain: sigma Hall 0} where both current vertices correspond to the corrections $\delta\hat{J}_\mu^\omega(\ep,\bk)$ and $\delta\hat{J}_\nu^\omega(\ep,\bk)$, see \eqn{explain: explicit delta J omega}, vanishes identically, 
in agreement with the results in Sec.~\ref{A toy model calculation}. 
We proceed without making use of this observation until that will become necessary.
\\
Under $\mu\leftrightarrow\nu$ and $\ell\leftrightarrow m$, we find that 
\beal
&\sigma^\text{1H}_{12} = -\fract{e^2}{2V}\,\sum_\bk\,T\sum_\ep\,\ep_{\mu\nu}\,\sum_{\ell \not= m}\,
\partial_\ep\,\mathcal{G}_\ell(i\ep,\bk)\,\mathcal{G}_m(i\ep,\bk)\;J^\omega_{\mu,\ell m}(\ep,\bk)\;
J^\omega_{\nu,m \ell}(\ep,\bk)\\
&\quad = -\fract{e^2}{4V}\,\sum_\bk\,T\sum_\ep\,\ep_{\mu\nu}\,\sum_{\ell \not= m}\,
\partial_\ep\,S_{\ell m}(\ep,\bk)\;\bra{\partial_\mu u_\ell(\ep,\bk)} u_m(\ep,\bk)\rangle
\langle u_m(\ep,\bk)\ket{\partial_\nu u_\ell(\ep,\bk)}\,,
\label{explain: sigma Hall 1H}
\eal
where \cite{Andrea-PRB2023}
\beal
S_{\ell m}(\ep,\bk) &= 2\ln\fract{\;i\ep-\ep_\ell(\ep,\bk)\;}{\;i\ep-\ep_m(\ep,\bk)\;}
-\fract{\;i\ep-\ep_\ell(\ep,\bk)\;}{\;i\ep-\ep_m(\ep,\bk)\;}
+\fract{\;i\ep-\ep_m(\ep,\bk)\;}{\;i\ep-\ep_\ell(\ep,\bk)\;}\\
&= 2i\,\text{arg}\ln\fract{\;i\ep-\ep_\ell(\ep,\bk)\;}{\;i\ep-\ep_m(\ep,\bk)\;}+ 2\ln\left|\fract{\;i\ep-\ep_\ell(\ep,\bk)\;}{\;i\ep-\ep_m(\ep,\bk)\;}\right|
+ K_{\ell m}(\ep,\bk)
\,.
\label{explain: S-1}
\eal 
Only the imaginary part of $S_{\ell m}(\ep,\bk)$, which is odd in $\ep$,  
contributes to $\sigma^\text{1H}_{12}(\bk)$ in \eqn{explain: sigma Hall 1H}. Moreover, 
the argument of the logarithm is discontinuous crossing $\ep=0$ if 
$\ep_\ell(0,\bk)\,\ep_m(0,\bk)=\ep_\ell(\bk)\,\ep_m(\bk)<0$, specifically,
\beal
\text{arg}\ln\fract{\;i\ep-\ep_\ell(\ep,\bk)\;}{\;i\ep-\ep_m(\ep,\bk)\;} &
\xrightarrow[\ep\to 0]{}\, -\pi\,
\text{sign}(\ep)\,\Big[\theta\big(-\ep_\ell(\bk)\big)
-\theta\big(-\ep_m(\bk)\big)\Big]\,.
\label{explain: S-2}
\eal
We hence write
\beal
\partial_\ep\,S_{\ell m}(\ep,\bk) &\simeq  4\pi\,i\,\delta(\ep)\,
\Big[\theta\big(-\ep_\ell(\bk)\big)
-\theta\big(-\ep_m(\bk)\big)\Big]
+ \partial_\ep\,K_{\ell m}(\ep,\bk)\,,
\label{explain: S-3}
\eal
dropping terms that are either real or odd in $\ep$. In the limit of vanishing temperature, we 
find that, see \eqn{explain: current w-limit ep=0},
\beal
\sigma^\text{1H}_{12} &= i\,\fract{e^2}{2V}\, \sum_\bk\,
\sum_{\ell\not=m}\,\ep_{\mu\nu}\;\fract{\;\theta\big(-\ep_\ell(\bk)\big)
-\theta\big(-\ep_m(\bk)\big)\;}{\big(\ep_\ell(\ep,\bk)-\ep_m(\ep,\bk)\big)^2}\;
J^\omega_{\mu,\ell m}(\bk)\,J^\omega_{\nu,m\ell}(\bk)\\
&\; - \fract{e^2}{4V}\,\sum_\bk\,\int\fract{d\ep}{2\pi}\,\ep_{\mu\nu}\,\sum_{\ell \not= m}\,
\partial_\ep\,K_{\ell m}(\ep,\bk)\;\bra{\partial_\mu u_\ell(\ep,\bk)} u_m(\ep,\bk)\rangle
\langle u_m(\ep,\bk)\ket{\partial_\nu u_\ell(\ep,\bk)}\\
&= \sigma^\text{H}_{0\,12} + \sigma^{\ep \text{H}}_{12}\,,
\label{explain: sigma Hall 1H bis} 
\eal
where $\sigma^\text{H}_{0\,12}$ coincides with \eqn{Hall conductivity 1}.
The contribution $\sigma^\text{2H}_{12}$ in \eqn{explain: sigma Hall 0} can be 
instead written as 
\beal
\sigma^\text{2H}_{12} &= \fract{e^2}{2V}\,\sum_\bk\int\fract{d\ep}{2\pi}\,\ep_{\mu\nu}\sum_{\ell m n}{^{'}}
\mathcal{G}_\ell(i\ep,\bk)\,\mathcal{G}_m(i\ep,\bk)\,\mathcal{G}_n(i\ep,\bk)\,J^\omega_{\mu,\ell m}(\ep,\bk)\,F^\ep_{mn}(\ep,\bk)\,J^\omega_{\nu,n \ell}(\ep,\bk)\\
&\quad -\fract{e^2}{4V}\,\sum_\bk\int\fract{d\ep}{2\pi}\,\ep_{\mu\nu}\sum_{\ell\not=m}\,\partial_\mu\,K_{\ell m}(\ep,\bk)\;
\bra{\partial_\ep u_\ell(\ep,\bk)} u_m(\ep,\bk)\rangle\langle u_m(\ep,\bk)
\ket{\partial_\nu u_\ell(\ep,\bk)}\\
&\quad -\fract{e^2}{4V}\,\sum_\bk\int\fract{d\ep}{2\pi}\,\ep_{\mu\nu}\sum_{\ell\not=m}\,\partial_\nu\,K_{\ell m}(\ep,\bk)\;
\bra{\partial_\mu u_\ell(\ep,\bk)} u_m(\ep,\bk)\rangle\langle u_m(\ep,\bk)
\ket{\partial_\ep u_\ell(\ep,\bk)}\\
&= \sigma^\text{3H}_{12} + \sigma^{\mu \text{H}}_{12}+ \sigma^{\nu \text{H}}_{12}
\,,
\label{explain: sigma Hall 2H}
\eal
where the apex on the sum means that $\ell$, $m$ and $n$ are all different, 
and we replaced $S_{\ell m}(\ep,\bk)$ with $K_{\ell m}(\ep,\bk)$ since the 
singular term in $\ep$ does not contribute.\\
One can readily show, see \cite{Andrea-PRB2023} for details, that, since 
\bealn
\mathcal{G}_\ell(i\ep,\bk)\,\mathcal{G}_m(i\ep,\bk)\,\mathcal{G}_n(i\ep,\bk) &= 
\fract{K_{\ell m}(\ep,\bk)+K_{m n}(\ep,\bk)+K_{n \ell }(\ep,\bk)}
{\big(\ep_\ell(\ep,\bk)-\ep_m(\ep,\bk)\big)\big(\ep_m(\ep,\bk)-\ep_n(\ep,\bk)\big)
\big(\ep_n(\ep,\bk)-\ep_\ell(\ep,\bk)\big)}\,,
\eal
then, recalling that $K_{\ell \ell}(\ep,\bk)=0$, 
\beal
\sigma^\text{3H}_{12} &= -\fract{e^2}{4V}\,\sum_\bk\int\fract{d\ep}{2\pi}\,\ep_{\mu\nu}\sum_{\ell m}\,K_{\ell m}(\ep,\bk)\,\Big\{\\
&\qquad\qquad\qquad\qquad\qquad \qquad
\partial_\mu\,\big(\bra{\partial_\ep u_\ell(\ep,\bk)}u_m(\ep,\bk)\rangle\langle 
u_m(\ep,\bk)\ket{\partial_\nu u_\ell(\ep,\bk)}\big)\\
&\qquad\qquad\qquad\qquad\qquad \qquad
+\partial_\ep\,\big(\bra{\partial_\nu u_\ell(\ep,\bk)}u_m(\ep,\bk)\rangle\langle 
u_m(\ep,\bk)\ket{\partial_\mu u_\ell(\ep,\bk)}\big)\\
&\qquad\qquad\qquad\qquad\qquad \qquad\quad
+\partial_\nu\,\big(\bra{\partial_\mu u_\ell(\ep,\bk)}u_m(\ep,\bk)\rangle\langle 
u_m(\ep,\bk)\ket{\partial_\ep u_\ell(\ep,\bk)}\big)\Big\}\,.
\label{explain: sigma 3H}
\eal
It follows that 
\beal
&\sigma^\text{3H}_{12}+\sigma^{\mu \text{H}}_{12} +\sigma^{\ep \text{H}}_{12}+ \sigma^{\nu \text{H}}_{12} = -\fract{e^2}{4V}\,\sum_\bk\int\fract{d\ep}{2\pi}\,\ep_{\mu\nu}\sum_{\ell m}\,\Big\{\\
&\qquad\qquad\qquad\qquad
\partial_\mu\,\big(K_{\ell m}(\ep,\bk)\,\bra{\partial_\ep u_\ell(\ep,\bk)}u_m(\ep,\bk)\rangle\langle 
u_m(\ep,\bk)\ket{\partial_\nu u_\ell(\ep,\bk)}\big)\\
&\qquad\qquad\qquad\qquad
+\partial_\ep\,\big(K_{\ell m}(\ep,\bk)\,\bra{\partial_\nu u_\ell(\ep,\bk)}u_m(\ep,\bk)\rangle\langle 
u_m(\ep,\bk)\ket{\partial_\mu u_\ell(\ep,\bk)}\big)\\
&\qquad\qquad\qquad\qquad\quad
+\partial_\nu\,\big(K_{\ell m}(\ep,\bk)\,\bra{\partial_\mu u_\ell(\ep,\bk)}u_m(\ep,\bk)\rangle\langle 
u_m(\ep,\bk)\ket{\partial_\ep u_\ell(\ep,\bk)}\big)\Big\}\,,
\label{explain: sigma 3H + all}
\eal
is the sum of full derivatives of continuous functions, periodic both in $\bk$ and $\ep$, 
noticing that at $\ep\to\pm \infty$ we recover the Hartree-Fock results. If both current vertices 
correspond to the unrenormalized ones, thus $\partial_\mu=\partial_{k_\mu}$ 
and $\partial_\nu=\partial_{k_\nu}$, \eqn{explain: sigma 3H + all} vanishes 
identically, consistently with the results of \cite{Andrea-PRB2023}. Therefore, 
the correction $\delta\sigma^\text{H}_{12}$ to $\sigma^\text{H}_{0\,12}$ in \eqn{explain: sigma Hall 1H bis} comes from the terms where $\partial_\mu=\partial_{\kappa_\mu}$, and thus 
$\partial_\nu=\partial_{k_\nu}$, or vice versa, which, being equal, 
imply, going back to \eqn{explain: sigma Hall 1H bis} and \eqn{explain: sigma Hall 2H} and through 
\eqn{explain: explicit delta J omega}, that 
\beal
\delta\sigma^\text{H}_{12} &= \fract{e^2}{V}\,\sum_\bk\int\fract{d\ep}{2\pi}\,\ep_{\mu\nu}\sum_{\ell m n}{^{'}}
\mathcal{G}_\ell(i\ep,\bk)\,\mathcal{G}_m(i\ep,\bk)\,\mathcal{G}_n(i\ep,\bk)
\,\bra{u_\ell(\ep,\bk)}\delta\hat{J}^\omega_\mu(\ep,\bk)\ket{u_m(\ep,\bk)}\\
&\qquad \qquad\qquad
\bra{u_m(\ep,\bk)}\partial_\ep\hat{H}_*(\ep,\bk)\ket{u_n(\ep,\bk)}\,\bra{u_n(\ep,\bk)}\partial_{k_\nu}\hat{H}_*(\ep,\bk)\ket{u_\ell(\ep,\bk)}
\\
&\;
- \fract{e^2}{2V}\,\sum_\bk\,\int\fract{d\ep}{2\pi}\,\ep_{\mu\nu}\,\sum_{\ell \not= m}\,
\partial_\ep\,K_{\ell m}(\ep,\bk)\;\bra{\delta_\mu u_\ell(\ep,\bk)} u_m(\ep,\bk)\rangle
\langle u_m(\ep,\bk)\ket{\partial_\nu u_\ell(\ep,\bk)}\\
&\; -\fract{e^2}{2V}\,\sum_\bk\int\fract{d\ep}{2\pi}\,\ep_{\mu\nu}\sum_{\ell\not=m}\,\delta_\mu K_{\ell m}(\ep,\bk)\;
\bra{\partial_\ep u_\ell(\ep,\bk)} u_m(\ep,\bk)\rangle\langle u_m(\ep,\bk)
\ket{\partial_{k_\nu} u_\ell(\ep,\bk)}\\
&\; -\fract{e^2}{2V}\,\sum_\bk\int\fract{d\ep}{2\pi}\,\ep_{\mu\nu}\sum_{\ell\not=m}\,\partial_{k_\nu} K_{\ell m}(\ep,\bk)\;
\bra{\delta_\mu u_\ell(\ep,\bk)} u_m(\ep,\bk)\rangle\langle u_m(\ep,\bk)
\ket{\partial_\ep u_\ell(\ep,\bk)}\,,
\label{explain: sigma Hall 2H disperazione}
\eal
where $\delta_\mu$ represents the first order expansion in 
$\delta\hat{J}^\omega_\mu(\ep,\bk)$ of eigenstates and eigenvalues of the Hamiltonian \eqn{explain: new Hamiltonian}, thus 
\bealn
\ep_\ell(\ep,\bk) &\to\; \ep_\ell(\ep,\bk)+\delta_\mu\ep_\ell(\ep,\bk)=
\ep_\ell(\ep,\bk)+\delta\hat{J}_{\mu,\ell\ell}^\omega(\ep,\bk)\,,\\
\ket{u_\ell(\ep,\bk)} &\to\; \ket{u_\ell(\ep,\bk)} +  
\ket{\delta_\mu u_\ell(\ep,\bk)} = \ket{u_\ell(\ep,\bk)} +
\sum_{m\not=\ell}\,\fract{\delta\hat{J}_{\mu,m\ell}^\omega(\ep,\bk)}
{\;\ep_\ell(\ep,\bk)-\ep_m(\ep,\bk)\;}\ket{u_m(\ep,\bk)}
\,,
\eal
and, we recall, 
\bealn
K_{\ell m}(\ep,\bk) &=\big(\ep_\ell(\ep,\bk)-\ep_m(\ep,\bk)\big)\,
\Big(\mathcal{G}_\ell(i\ep,\bk)+\mathcal{G}_m(i\ep,\bk)\Big)\,,
\eal
depends on the eigenvalues. 
We first note that the last two terms on the right hand side of \eqn{explain: sigma Hall 2H disperazione} involve the diagonal, intraband, matrix elements of the 
currents $\delta\hat{J}^\omega_\mu(\ep,\bk)$ and $\hat{J}^q_\nu(\ep,\bk)$, which cannot 
contribute to the anomalous Hall conductivity. Therefore, we can safely discard those two terms.\\
Concerning instead the first two terms on the right hand side of \eqn{explain: sigma Hall 2H disperazione}, we observe that they require, to be finite, 
phase coherence between the matrix elements connecting two or three distinct bands 
through different operators. Specifically, one of the matrix elements is that of 
$\delta\hat{J}^\omega_\mu(\ep,\bk)$ in 
\eqn{explain: explicit delta J omega}, which represents an interaction scattering 
amplitude between a quasiparticle-quasihole pair in bands $\ell$ and $m$ at finite frequency $\ep$, which has to be integrated, and an intraband pair at $\ep=0$ on the quasiparticle Fermi surface. 
It is conceivable, given the decoherence effects brought about by interaction, that such process, just because of the different frequencies of the two pairs, cannot  
maintain phase coherence upon integrating over frequencies and summing over all bands and momenta.  
On the contrary, phase coherence is maintained in $\sigma^\text{H}_{0\,12}$ of \eqn{explain: sigma Hall 1H bis}, where both pairs are at $\ep=0$. We shall therefore assume that the first two terms on the right hand side of \eqn{explain: sigma Hall 2H disperazione} averages at zero, 
thus leading to $\delta\sigma^\text{H}_{12}=0$ and to the desired result  
\be
\sigma^\text{H}_{12} = \sigma^\text{H}_{0\,12}= i\,\fract{e^2}{2V}\, \sum_\bk\,
\sum_{\ell\not=m}\,\ep_{\mu\nu}\;\fract{\;\theta\big(-\ep_\ell(\bk)\big)
-\theta\big(-\ep_m(\bk)\big)\;}{\big(\ep_\ell(\ep,\bk)-\ep_m(\ep,\bk)\big)^2}\;
J^\omega_{\mu,\ell m}(\bk)\,J^\omega_{\nu,m\ell}(\bk)\,.
\label{explain: sigma Hall 1H disperazione finale} 
\ee
Let us finally discuss the multiband extension of same results in Sec.~\ref{A toy model calculation} for the case $\ep=0$, where we believe phase coherence does hold.  
From the discussion in that section, considering two different states $\alpha\not=\beta$ 
in the original basis and the Pauli matrices  
$\sigma_1$ and $\sigma_2$ in this subspace, thus the off-diagonal 
generalised Pauli matrices, 
if 
\beal
H_{*,\alpha\beta}(0,\bk) &=\big|H_{*,\alpha\beta}(0,\bk)\big|\,\esp{-i\phi_{\alpha\beta}(\bk)}\\
&=\big|H_{*,\alpha\beta}(0,\bk)\big|\,
\big(\cos\phi_{\alpha\beta}(\bk)\,\sigma_{1,\alpha\beta}
+ \sin\phi_{\alpha\beta}(\bk)\,\sigma_{2,\alpha\beta}\big)\,,
\label{explain: H* alpha beta}
\eal
then we can write
\bealn
\delta J^\omega_{\mu,\alpha\beta}(0,\bk) 
&=\Rea\,\delta J^\omega_{\mu,\alpha\beta}(0,\bk)\,
\big(\cos\phi_{\alpha\beta}(\bk)\,\sigma_{1,\alpha\beta}
+ \sin\phi_{\alpha\beta}(\bk)\,\sigma_{2,\alpha\beta}\big)\,.
\eal 
Therefore, while  
\bealn
\partial_{k_\mu} H_{*,\alpha\beta}(0,\bk) &=
J^q_{\mu,\alpha\beta}(0,\bk)\\
&=
\partial_{k_\mu}\big|H_{*,\alpha\beta}(0,\bk)\big|\;
\big(\cos\phi_{\alpha\beta}(\bk)\,\sigma_{1,\alpha\beta}
+ \sin\phi_{\alpha\beta}(\bk)\,\sigma_{2,\alpha\beta}\big)\\
&\quad - \partial_{k_\mu}\phi_{\alpha\beta}(\bk)\;
\big|H_{*,\alpha\beta}(0,\bk)\big|\,
\big(\sin\phi_{\alpha\beta}(\bk)\,\sigma_{1,\alpha\beta}
- \cos\phi_{\alpha\beta}(\bk)\,\sigma_{2,\alpha\beta}\big)\,,
\eal
where the interference between the two terms is responsible for the possibly non-trivial topology, 
$\delta J^\omega_{\mu,\alpha\beta}(0,\bk)$ is instead proportional to only one component of $J^q_{\mu,\alpha\beta}(0,\bk)$, specifically,   
\bealn
\delta J^\omega_{\mu,\alpha\beta}(0,\bk) 
&\propto 
\partial_{k_\mu}\big|H_{*,\alpha\beta}(0,\bk)\big|\;
\big(\cos\phi_{\alpha\beta}(\bk)\,\sigma_{1,\alpha\beta}
+ \sin\phi_{\alpha\beta}(\bk)\,\sigma_{2,\alpha\beta}\big)\,,
\eal
and thus cannot contribute 
on its own to the anomalous Hall conductivity, which is a further justification of 
neglecting the second order term in $\delta\,\, \hat{\!\!\bd{J}}^\omega(\ep,\bk)$. \\
These properties survive in the diagonal basis, too. 
It follows that, if we write, for $\ell\not=m$ and using similar conventions as in 
Sec.~\ref{A toy model calculation} but in terms of generalized Pauli matrices,   
\beal
J^q_{\mu,\ell m}(0,\bk)&= J^q_{2\mu,\ell m}(0,\bk)
\big(\cos\phi_{\ell m}(\bk)\,\sigma_{1,\ell m}
+ \sin\phi_{\ell m}(\bk)\,\sigma_{2,\ell m}\big)\\
&\quad +J^q_{1\mu,\ell m}(0,\bk)\,
\big(\sin\phi_{\ell m}(\bk)\,\sigma_{1,\ell m}
- \cos\phi_{\ell m}(\bk)\,\sigma_{2,\ell m}\big)\\
&= \text{J}^q_{2\mu,\ell m}(\bk)
+ \text{J}^q_{1\mu,\ell m}(\bk)
\,,
\label{explain: assumption Jq}
\eal
with real $J^q_{1\mu,\ell m}(0,\bk)$ and $J^q_{2\mu,\ell m}(0,\bk)$, 
in which only the interference between the two may cause a non-trivial topology, 
then we are allowed to write $\delta J^\omega_{\mu,\ell m}(0,\bk)$ proportional to only one of them, e.g., and without loss of generality,  
\beal
\delta J^\omega_{\mu,\ell m}(0,\bk) &= F^1_{2\mu,\ell m}(\bk)\,
\text{J}^q_{2\mu,\ell m}(\bk)\,,
\label{explain: assumption Jw}
\eal
also with real $F^1_{2\mu,\ell m}(\bk)$. Moreover, if the Fermi surface is isotropic, as we here assume, 
$F^1_{2\mu,\ell m}(\bk)=F^1_{2,\ell m}(\bk)$ is independent of the direction $\mu$.\\

\noindent
In conclusion, we have shown that the anomalous Hall conductivity 
$\sigma^\text{H}_{12}$ in \eqn{explain: sigma Hall 1H disperazione finale} reduces to the desired Fermi-liquid expression \eqn{Hall conductivity 1}. In other words, we have demonstrated that also in a multi-band Fermi liquid, 
besides the already well-known single-band case \cite{RevModPhys.47.331,libro}, 
there is perfect coincidence between the physical transport coefficients 
and those obtained through the interacting Hamiltonian \eqn{Hqp}
treated within Hartree-Fock plus RPA. Such correspondence holds on condition that the Hartree-Fock Hamiltonian \eqn{self-consistency 1} coincides with 
$\hat{H}_*(\ep,\bk)$ in \eqn{H*(epsilon)} calculated at $\ep=0$, and that 
the RPA interaction vertices in the dynamic and static limits are equal to 
the quasiparticle ones, $\hat{\Gamma}^\omega_*(0,\bk;0,\bkp)$  
and $\hat{\Gamma}^q_*(0,\bk;0,\bkp)$, respectively. The additional assumption that we make 
besides the standard Fermi liquid ones is that a scattering process between distinct bands induced by   the dynamic correction $\delta\,\, \hat{\!\!\bd{J}}^\omega(\ep,\bk)$ to the current does not maintain phase coherence once summed over frequency, momentum and band indices, thus averaging at zero. 
The only exceptions are quantities that catch the singularities at zero frequency which arise from the 
discontinuity of the Green's function phase at $\ep=0$, see, e.g., \eqn{explain: S-3}, 
since $\delta\,\, \hat{\!\!\bd{J}}^\omega(0,\bk)$ does describe a phase coherent process.


\bibliography{mybiblio}
\end{appendix}




\nolinenumbers

\end{document}